\documentstyle[12pt,epsf]{article}

\def\bar{\overline}          
          
\def\a{\alpha}  
\def\b{\beta}

\def\bar{\overline}

\def\be{\begin{equation}}  
\def\ee{\end{equation}}  
\def\bea{\begin{eqnarray}}  
\def\eea{\end{eqnarray}}         
\def\D{\Delta}
\def\dmsq{\Delta m^2}

\begin{document}          
\setcounter{page}{0}          
\thispagestyle{empty}                          
\topskip 0. cm 
\voffset = -.60 in 
\hoffset = -.50 in 
\begin{flushright}      
CERN-TH/2001-141 \\    
FERMILAB-Pub-01/078-T \\
MPI-PhT/2001-15\\
Lund-Mph-01/02 
\end{flushright}          
\vspace{0.3 cm}          
\centerline{\Large\bf Combining LSND and Atmospheric Anomalies} 
\vspace{0.2cm}
\centerline{\Large\bf in a Three-Neutrino Picture}  
\vskip 1 cm          
\begin{center} 
{\bf Gabriela Barenboim 
\footnote{E-mail address: gabriela@fnal.gov}${}^{a,b}$, 
Amol Dighe
\footnote{E-mail address: amol@mppmu.mpg.de}${}^{a,c}$}   
and 
{\bf Solveig Skadhauge 
\footnote{E-mail address: Solveig.Skadhauge@matfys.lth.se}${}^{d}$} 
\end{center}
\centerline{${}^a$ \it{CERN - TH Division, CH-1211 Geneva 23,
Switzerland.}}  
\centerline{${}^b$ \it{FERMILAB Theory Group, MS 106, P.O.Box 500, 
Batavia IL 60510, USA.}} 
\centerline{${}^c$ \it{Max-Planck-Institut f{\"u}r Physik, 
F{\"o}hringer Ring 6, D-80805 Munich, Germany.}}
\centerline{${}^d$ \it{Department of Mathematical Physics, LTH, 
Lund University, S-22100 Lund, Sweden.}} 
\bigskip
\centerline{May 2001}
\bigskip         
\centerline{\bf ABSTRACT}\par       
\vskip 0.5 cm   
We investigate the three-neutrino mixing scheme for solving the 
atmospheric and LSND anomalies.
We find the region in the parameter space that provides a 
good fit to the LSND and the SK atmospheric data, taking into
account the CHOOZ constraint.
We demonstrate that the goodness of this fit is comparable to
that of the conventional fit to the solar and atmospheric data. 
Large values of the LSND angle are favoured and 
$\sin^2(2\theta_{\rm LSND})$ can be as high as 0.1. 
This can have important effects on the atmospheric 
electron neutrino ratios as well as on down-going multi-GeV 
muon neutrino ratios. 
We examine the possibility of distinguishing this scheme from the
conventional one at the long baseline experiments.
We find that the number of electron neutrino events observed
at the CERN to Gran Sasso experiment may lead us to identify
the scheme, and hence the mass pattern of neutrinos.

\newpage 


\section{Introduction}
\label{intro}

The present data from the experiments on atmospheric, solar and 
accelerator (LSND) neutrinos indicate neutrino flavour oscillations. 
The data from each of these sets of experiments individually
can be explained by a single dominant mass square difference
$\dmsq$ and a mixing angle $\theta$ between two active neutrinos.
The atmospheric neutrino data from SuperKamiokande (SK) \cite{SKevi}
indicate $\nu_\mu \leftrightarrow \nu_\tau$ as the dominant mode,
with 
$\Delta m^2_{\rm atm} = (1$ - $8) \times 10^{-3} \mbox{ eV}^2, 
\sin^2 2\theta = 0.8$ - $1.0$.
The three MSW solutions (LMA, SMA and LOW) as well as the vacuum 
oscillations can provide reasonable fits to the solar neutrino data 
\cite{Sksolar,homestake,sage,gallex} and all these solutions have
$\dmsq_\odot \leq 2 \times 10^{-4}$ eV$^2$. 
The results of the LSND experiment \cite{LSND,lsndnew} are neither
confirmed nor fully excluded by the KARMEN2 data \cite{karmen}, 
and the combined fit allows a region \cite{lsnd-kar} of
$\Delta m^2_{LSND} = (0.1$ - $1) \mbox{ eV}^2, 
\sin^2 2\theta_{\mu e} \approx 10^{-3} -  10^{-2}$.
There are also constraints on the mixing of $\nu_e$ from the CHOOZ 
experiment \cite{chooz}: we have $\sin^2 2 \theta_e \leq 0.1$ 
for $\dmsq > 10^{-3}$ eV$^2$.

In the context of only three known neutrino species, 
the $\dmsq$s corresponding to the solutions of the three
neutrino anomalies above (atmospheric, solar and LSND) cannot
be reconciled. The three-neutrino schemes make a very poor
fit to all the data. Two ways are possible out of this predicament:
(i) turn a blind eye to one of the experiments and fit for the
other two in three-neutrino schemes;
(ii) solve two of the neutrino anomalies with three-neutrino 
oscillations and solve the third one by using
exotic models such as sterile neutrinos, FCNC, neutrino decay,
extra dimensions, etc.

Since the LSND result is yet to be confirmed, it is customary 
to ignore it and accommodate
the solar neutrino deficit and the atmospheric anomaly
with the mixing between three active neutrinos. Implicit in 
this is the assumption that the LSND results will be proved 
false by future experiments, which can be justified by the 
fact that KARMEN2 \cite{karmen} and Bugey \cite{bugey} already rule out 
most of the allowed region of LSND. 
However, this is just a convenient assumption,
and the possibility of the LSND results being confirmed by 
future experiments such as BooNe \cite{boone} cannot be ignored. 
Also the new analysis of the final data by the LSND 
collaboration \cite{lsndnew} is consistent with the 
old results \cite{LSND}, and therefore strengthens the anomaly 
evidence.

Our approach will be to study the neutrino 
anomalies with three-(active)-neutrino oscillation. 
This allows us to solve only two of the three anomalies.  
As the atmospheric data are showing strong evidence for 
neutrino oscillation, thanks to the large range of L/E probed, 
we will take this to be one of the anomalies solved by
oscillation. 
There is no compelling evidence that the electron neutrinos
participate in the oscillations of atmospheric neutrinos. This implies 
that the $P_{\nu_e \rightarrow \nu_\mu}$ must be small, 
meaning either that the mixing 
angle is small (LSND case) or that the $\dmsq$ is too small to affect the 
atmospheric neutrinos (solar case). The large angle solutions to LSND 
is in any case ruled out by the results of Bugey. 
Moreover, just from the point of view of goodness of fit
(quantified by a $\chi^2$ function), the best fits to 
(I) atmospheric and solar data, and (II) atmospheric and LSND
data are equally good (as we shall show in this paper). 
Scheme II gives a different mass spectrum from the conventional
scheme I, and an eye should be kept on future experiments 
in order to resolve this discrete ambiguity.
That we leave out one anomaly does not mean that we do not believe in
it, but rather that it has to be solved in some other way.  

In the case of solar neutrinos there are various viable ``exotic''
solutions. The most popular among them is the sterile neutrino SMA MSW
solution. In the favoured ``2+2'' four-neutrino scheme \cite{22},
the sterile neutrino participates mainly in the solar neutrino anomaly 
and the three active neutrinos solve the atmospheric and LSND 
anomalies among themselves \cite{sterile}. This is indeed a possible 
extension of
scheme II, and the fit we perform will be a guide for this option.
Many other exotic solutions exist for solving the solar neutrino
anomaly \cite{solarexotic}.  
Flavour changing neutral current interactions (FCNC) \cite{solarfcnc} 
can explain the data well \cite{fcncfit} --
this solution uses the matter of the Sun to make a FCNC 
transition, such as $\nu_e$ + quark $\rightarrow$ quark + $\nu_\mu$. 
The different neutrino production points gives the necessary energy 
dependence of the solar neutrino flux. Furthermore a violation of 
the equivalence principle has been suggested and cannot be 
excluded. Solutions have also been suggested in the scenarios with
large extra dimensions, where the solar neutrino anomaly is
accounted for by the mixing with a Kaluza-Klein tower of
sterile neutrinos \cite{ds} and the three active neutrinos 
account for the atmospheric and LSND anomalies \cite{dj}.
A resonant spin-flip conversion \cite{solarmag} of the 
left-handed neutrino to unobservable right-handed states, due to the 
solar magnetic field, gives a good fit to the data \cite{magfit}. 
This solution cannot be reconciled with LSND and atmospheric anomalies, 
as the mass squared difference needed is too small 
($\approx 10^{-8}$eV$^2$). 
In the three-neutrino schemes able to account for the 
atmospheric and LSND result, one of the possible mass patterns 
has the electron neutrino mainly in the heavy state, and
therefore neutrino decay could also be thought of as a way out for  
explaining the solar anomaly. Although this solution does not agree 
well with the data.

Some exotic solutions for the LSND anomaly have also 
been suggested in the literature \cite{lsndex}: for instance, 
having new flavour-violating decay modes of muons. 
This possibility cannot yet be ruled out in a model-independent way.  
The introduction of a sterile neutrino to solve the LSND anomaly
gives rise to the ``3+1'' neutrino mass pattern \cite{31},
which is very close to being excluded \cite{31and22}.
It might be argued that the exotic solutions are more likely
to solve the solar anomaly than the LSND one, since
the matter and the magnetic field, which provide more degrees of
freedom for the exotic solutions in the solar case,
are absent in the LSND experiment.

There is a vast literature on three-neutrino oscillation fits 
to atmospheric data \cite{threefit}.  
Also the combined fit to solar and atmospheric data is well-explored 
and is known to describe the data well \cite{gonz,glothree}. 
Furthermore the two-neutrino 
$\nu_\mu \rightarrow \nu_\tau$ oscillation gives a good explanation of the 
atmospheric data. This solution would be obtained in the limit of a very small 
LSND angle, and therefore we already know that the LSND and atmospheric data 
can be reconciled. The important question to answer is if the 
three-neutrino scheme can fit the data even better and if there are 
any possible effects that could be observed in the near future.

The three-active-neutrino solution to LSND and atmospheric 
problems has also been studied in \cite{choubey}, where however 
only the up/down asymmetries from the atmospheric data 
were used for the fits. Here we will consider the full set of 40 
data points from the sub-GeV and 
multi-GeV neutrinos observed by the SK collaboration. Furthermore 
we take into account the data from CHOOZ \cite {chooz}
and the matter effects inside the Earth.   
We define a $\chi^2$ function and perform 
fits to schemes I and II above, and find that both fits are
equally good. We explore further the fit to scheme II. 
The mass pattern corresponding to this fit may be tested at the
long baseline experiments, e.g. K2K \cite{k2k}, MINOS \cite{minos}
or CNGS \cite{cgs}. 
We perform Monte Carlo simulations
to check if this mass pattern can be distinguished from the 
conventional one at these experiments.
If BooNE confirms the LSND results, 
this scheme II, coupled with the appropriate exotic solution
for the solar neutrino anomaly,
will provide the solution for the mass spectrum of
neutrinos.

The paper is organized as follows. In Sec.~\ref{fits} we perform
fits to the data for schemes I and II.
In Sec.~\ref{compare}, 
we compare the signals at the long baseline
experiments with the two mass patterns and look for ways of
distinguishing between them.
Our findings are summarized in Sec.~\ref{concl}.


\section{$\chi^2$ fits to the data}
\label{fits}

The aim of this section is to find the regions in the three-neutrino
mixing parameter space that describe (I) the atmospheric and solar data
and (II) the atmospheric and LSND data. 
We shall define a $\chi^2$ function in order to quantify the
goodness of these fits. It is clearly not fair to compare the
$\chi^2$s from these two fits, since they correspond to different data
sets. We perform the fit to scheme I to demonstrate the soundness of
our $\chi^2$ function by showing that this procedure, rough though it is,
reproduces the conventional fits, which are much more accurate
\cite{threefit,gonz}. Since our aim here is just to get a broad estimate of
the allowed parameter space, this rough fit should suffice for
our purpose. Then we follow the same procedure to perform a fit
to scheme II and find the allowed region in the parameter space that 
describes the data reasonably well.

The neutrino mixing matrix $U_{MNS}$ 
is parametrized as \cite{pdg}
\be
  U_{MNS}=\left(\matrix{ c_{12} c_{13} & s_{12}c_{13} & s_{13} \cr
-s_{12}c_{23}-c_{12}s_{23}s_{13} & c_{12}c_{23}-s_{12}s_{23}s_{13} 
& s_{23}c_{13} \cr
s_{12}s_{23}-c_{12}c_{23}s_{13} & -c_{12}s_{23}-s_{12}c_{23}s_{13} &
c_{23}c_{13} }\right)
\label{u-mns}
\ee
where we have neglected any $CP$ violation. The mass spectrum is 
parametrized by five parameters,  $s_{12}$, $s_{23}$, $s_{13}$, 
$\Delta m_{21}^2$ and $\Delta m_{31}^2$. (Here and in the following 
$s_{ij}$, $c_{ij}$ are shorthands for $\sin(\theta_{ij})$, 
$\cos(\theta_{ij})$ respectively, and we use the convention
$\Delta m^2_{ij} \equiv m^2_i - m^2_j$.)
We define $\nu_1$ as the lightest state and $\nu_3$ as 
the heaviest state. 

For both schemes there are two solutions; the normal hierarchy 
($\dmsq_{21} \ll \dmsq_{32} \simeq \dmsq_{31}$) and the 
inverted hierarchy ($ \dmsq_{31} \simeq \dmsq_{21} \gg \dmsq_{32}$).
Neutrino oscillations in vacuum cannot distinguish between the two
hierarchies, but with matter effects, the data could in principle
distinguish between them.
In scheme I the characteristic features of the normal hierarchy is the small 
value of $s_{13}$, and the inverted hierarchy corresponds to values of 
$s_{12}$ close to 1. 
It has recently been shown that the differences between the two hierarchies 
within this scheme are very small \cite{newfogli}. 
For scheme II normal (inverted) hierarchy corresponds to  
$s_{13}\simeq 1$ ($s_{13},s_{12} \simeq 0$). 
The matter effects in this scheme are negligible, since
both the $\dmsq_{\rm atm}$ and $\dmsq_{\rm LSND}$ are too large for
the Earth's density to play any significant part. Therefore,
the differences between the predictions from the two hierarchies 
are extremely small. 
For both schemes we shall only consider the normal hierarchy while 
performing the fit.
It should be remembered that for each point in the parameter space
with normal hierarchy, there exists a corresponding point with inverted
hierarchy that gives almost the same value of $\chi^2$.

For the atmospheric data, we only consider the 
SK multi-GeV and sub-GeV data \cite{SKnu2000}.
The mean neutrino energy for the partially contained events 
and the upward going muons at SK is higher. In this range the effects of 
a solar mass squared difference are therefore negligible. 
The small contributions arising from an averaging of a LSND mass sqaured 
difference can be compensated by a shift towards smaller $\dmsq_{\rm atm}$ 
(see section \ref{atm-lsnd}). Therefore it is not expected that the data 
will affect the comparison of the two schemes. Nevertheless it might 
results in small changes in the allowed regions.  
The further inclusion of the data from
other atmospheric neutrino experiments \cite{otheratm}
is not expected to affect the fit much, since the number of fully contained 
events at SK is overwhelmingly large as compared to these.

The experimental data are represented by the ratios $R_{\alpha,i}$ 
between the experimental values for the fluxes and the 
theoretical Monte Carlo prediction in the case of no oscillation 
for muon and electron neutrinos in the 10 different zenith angle bins.  
The ratios $R_{\alpha,i}$ can be written as
\be
  R_{\alpha,i}^{\rm exp}= N_{\alpha,i}^{\rm exp}/N_{\alpha,i}^{\rm MC}
  \;\;\;\;,\;\a =\mu,e \;\;,\;i=1\dots10 \;.
\ee

The number of events for the sub-GeV neutrinos is calculated as;
\bea
  N_{\alpha} &=& n_T \sum_{\nu, \bar \nu}\sum_{\beta=e,\mu} 
       \int_0^{\infty}dE_{\nu} 
       \int_{-1}^{1}d\cos(\theta_{\nu}) 
        \int_{E_{l,{\rm min}}}^{E_{l,{\rm max}}}
        dE_l \int_{-1}^{1} d \cos(\theta_{l\nu}) 
       \frac{1}{2\pi}\int_0^{2\pi} d\phi 
 \nonumber \\ 
     & \times & \frac{d^2\Phi_{\beta}}{dE_{\nu} d\cos(\theta_\nu)} 
        \frac{d^2\sigma_{\beta}}{dE_ld\cos(\theta_{l\nu})}
        P_{\alpha \beta}(E_{\nu},\theta_\nu ) \epsilon (E_l) \;,  
\eea
where $E_{\nu}$ is the neutrino energy, $E_l$ is the lepton energy, 
$\theta_{l\nu}$ is the angle between the neutrino and the 
scattered lepton, $\theta_{\nu}$ is the zenith angle of neutrino. 
The number of target nucleons is denoted by $n_T$. 
$\phi$ is the azimuthal angle of the incoming neutrino, and is used to 
calculate the zenith angle $\theta_l$ of the charged lepton:
\be
  \cos(\theta_{l})= \cos(\theta_{\nu})\cos(\theta_{\nu l})+\sin(\theta_{\nu})
                      \cos(\phi)\sin(\theta_l) \;. 
\ee
For the differential sub-GeV cross section for $\nu_l N \rightarrow l X$, 
we use the quasi-elastic 
approximation \cite{subcross}, with the sub-GeV fluxes $\Phi$ 
taken from \cite{subflux}.  
The efficiency function $\epsilon(E_l)$ is not 
published by the SK collaboration. But as we divide by the SK Monte 
Carlos, this effect is averaged out if the efficiency 
function is flat within the different samples.

In the calculations of the multi-GeV ratios 
we find that the number of events can be well described by 
\be\label{mulne}
  N_l = n_T \int_{E_{l,min}}^{E_{l,max}} dE_{\nu} 
       \int_{-1}^{1}d\cos(\theta_{\nu}) 
       \frac{d^2\Phi}{dE_{\nu} d\cos(\theta_{\nu})} 
       \sigma_{\alpha}(E_\nu) P_{\alpha \beta}(E_{\nu},\theta_{\nu} )   \;. 
\ee
Here we assume that the energy of the charged 
lepton is the same as that of the incoming neutrino. In order 
to account for the scattering angle, we further smear the 
spectrum with a Gaussian function, where we use the width 
$\sigma \approx 25^\circ$ for the electron neutrinos and
$\sigma \approx 15^\circ$ for the muons neutrinos \cite{smearing}. 
The total cross section for neutrinos is given in \cite{mulcross}, 
and for the antineutrinos we use the approximate formula 
$\sigma_{\bar{\nu}_e}=0.34 \times 10^{-38}E_\nu$ cm$^2$/GeV.
The multi-GeV fluxes are taken from 
\cite{mulflux}. A comparison between the ratios obtained from 
(\ref{mulne}) and the ones obtained by using the differential 
cross section including the quasi-elastic, one pion and deep inelastic 
channels yields hardly any difference. 

The probabilities are calculated taking 
into account all the mass-squared differences: 
\be
 P_{\a \b}= \delta_{\a \b} -4\sum_{i>j}U_{\a i} U_{\b j} U_{\a j} U_{\b i} 
\sin^2\left(\frac{\D m^2_{ij} L}{4E}\right) \;,
\ee
where all the quantities are calculated in the presence of matter
wherever appropriate.
Even in the conventional case, the sub-GeV ratios in particular 
are influenced also by the small mass squared difference \cite{dmsq}. 
The oscillation length $L$ depends on the zenith angle of the neutrino:
\be
  L= \sqrt{(R_E+h)^2-R_E^2\sin^2(\theta_{\nu})}-R_E\cos(\theta_\nu) \;,
\ee
where $R_E$ is the radius of the Earth and $h$ is the production 
height of the neutrinos in the atmosphere. We take $h\simeq 10$ km.

Matter effects are important for the atmospheric neutrinos. We simulate 
the matter effects by using a two-shell model of the matter densities in 
Earth. The density in the mantle (core) is taken to be roughly 
3.35 (8.44) g/cm$^3$, and the core radius is taken to be 2887 km.
From this we calculate the average density as a function of the 
neutrino zenith angle. This allows us to use the three-neutrino mixing 
matrix in matter, as calculated in \cite{threenum}.

We define the atmospheric $\chi^2$ as
\be
  \chi^2_{\rm atm}= \sum_{M, S}\sum_{\alpha=e,\mu}\sum_{i=1}^{10} 
     \frac{(R_{\alpha,i}^{\rm exp}-
     R_{\alpha,i}^{\rm th})^2}{\sigma_{\alpha i}^2} \quad ,
\ee
where $\sigma_{\alpha,i}$ are the statistical errors 
and $M,S$ stand for the multi-GeV and sub-GeV data respectively. 
The total number of data points is 40.

From the CHOOZ experiment \cite{chooz} we use the 15 data points 
with the statistical errors. The CHOOZ baseline is roughly 
1000 m, and the neutrino energy in the range from 3 to 9 MeV. Therefore the 
$\sin^2(\Delta m^2 L/4E)$ terms do oscillate within the energy region and  
hence all the data points need to be taken into account.
In each bin we average the probability over energy.
When including the LSND experiment we use one datum \cite{LSND}. The 
distance travelled by the antineutrinos ($\bar \nu_\mu$) is set to 30 m 
and we use the mean energy of 42 MeV. 
When including data from the Bugey experiment \cite{bugey} we take three 
data points; $P_{\bar \nu_e \bar \nu_e}$ for the three different baselines 
of 15 m, 40 m, and 95 m. The probability is averaged over the positron 
energy range 1 - 6 MeV. We take both the statistical and systematic errors, 
as the systematic errors are large compared to the statistical ones.

The solution to the solar neutrino 
problem is preferred to be within the LMA region and we will 
therefore use parameters only within this region in order to make a fit.
The $\chi^2$ for the solar LMA solution is taken from Ref.\cite{gonz}. This 
fit includes the total rates in the chlorine (Homestake)  experiment
\cite{homestake}, the gallium (SAGE, GALLEX) experiments 
\cite{sage,gallex}, Kamiokande \cite{solarkamio} and SuperKamiokande 
\cite{Sksolar}. 
Furthermore the spectra for both day and night are included.  
In total there are 42 data points fitting with three parameters: 
$\dmsq_{\odot}, \theta_{\odot}$ and the overall flux normalization 
for the SK spectrum.

The total $\chi^2$ function for scheme I is then
\be \label{chi1}
   \chi^2_{I}=\chi^2_{\rm atm}+\chi^2_{\rm CHOOZ}+\chi^2_{\rm LMA}~,
\ee
and for scheme II, we have
\be \label{chi2}
   \chi^2_{II}=\chi^2_{\rm atm}+\chi^2_{\rm CHOOZ}+\chi^2_{\rm LSND}~~.
\ee

Before interpreting the results, a word of caution is
in order. In our analysis (as in the standard procedures),  
we deal not with the number of observed events but with the ratio  
of the number of observed events to the number of events expected from 
the Monte Carlo. These ratios are convenient because they directly give a 
measure of the survival/oscillation probability. 
The Monte Carlo 
predictions, in the case of atmospheric neutrinos fluxes, contain
large uncertainties, especially in the absolute values of fluxes. 
This is the reason why some data are presented 
in the form of a ratio of two measured quantities and is compared with   
theoretical predictions of the same ratio. 
We have not included theoretical uncertainties and correlations in the
atmospheric neutrino fluxes. However, a previous analysis \cite{pilar} 
has found that they do not affect the fit significantly.

It has been pointed out recently that there is a 
significant discrepancy between the commonly used primary cosmic ray
fluxes and the measured ones \cite{flux}. The variation of the 
primary cosmic ray flux is directly related to the absolute value
of the atmospheric neutrino fluxes and can also affect
(to a less significant extent) its energy and zenith angle dependence. 
This may affect the allowed region of the atmospheric neutrino
anomaly. 
It is important to bear in mind that when applying the calculated 
atmospheric neutrino fluxes to a neutrino oscillation study, the
absolute values of the fluxes as well as the  tiny variations
with energy between them could become important.

\subsection{Fit to solar, atmospheric and CHOOZ data (scheme I)}
\label{atm-sol}

In this section, we perform a fit to the conventional three-neutrino
scheme for solving the solar and atmospheric anomalies.
When fitting within  this scheme we will take 
$\D m_{31}^2 \simeq 10^{-3}$ - $10^{-2}$ eV$^2$ and 
$\D m_{21}^2$ suitable to solve the solar neutrino problem. 
The $\chi^2_{\rm atm}$ and $\chi^2_{\rm CHOOZ}$ are to a very large extent
independent of $s_{12}$. We will therefore keep the solar angle  
constant at $s_{12}^2=0.3$, which is close to the best-fit point of 
the LMA solution \cite{gonz}.
The small $\Delta m_{21}^2$ is taken in the range 
$2 \times 10^{-5}$ - $2 \times 10^{-4}$ eV$^2$. 
In particular the electron neutrino ratios can be affected 
by the solar mass difference \cite{dmsq}, 
and we therefore calculate $\chi^2_{\rm atm}+\chi^2_{\rm CHOOZ}$ 
for a few values of $\dmsq_{\odot}$. 
We thus have four degrees of freedom,
$s_{23}, s_{13}, \dmsq_{21}, \dmsq_{31}$.

We first find the best point while doing a combined fit to 
SK and CHOOZ, with $\chi^2 \equiv \chi^2_{\rm atm} + \chi^2_{\rm CHOOZ}$. 
There are 55 data points: 40 from SK atmospheric data and 15 from
CHOOZ.
The minimum is  $\chi^2_{\rm min}=45.3$ at\footnote{ Henceforth,
we implicitly assume the units eV$^2$ for $\dmsq$, unless 
specified explicitly.} 
\be
s_{23}^2=0.41 \;,\; s_{13}^2=0.00 \;,\; 
\Delta m_{21}^2=2.0 \times 10^{-4} \;,\; 
\Delta m_{31}^2=4.1 \times 10^{-3}~~.
\ee
Note that the minimum occurs for the largest value in
the allowed region for $\dmsq_\odot$. 
With $\Delta m_{21}^2$ constrained to be $2.0 \times 10^{-5}$
($8.0 \times 10^{-5}$), the value of $\chi^2_{\rm min}$ is
$\chi^2_{\rm min}=49.1\; (46.9)$ at 
$s_{23}^2=0.58 \;(0.40),\:s_{13}^2=0.01 \;(0.00),\: 
\Delta m_{31}^2=0.0040 \;(0.0045)$. We see that the dependence on 
the solar mass squared difference is weak. The $\chi^2$ rises slightly 
when $\dmsq_\odot$ becomes smaller. 
For $\dmsq_\odot < 10^{-5}$ this dependence is 
lost and the $\chi^2$ function is almost constant.

\begin{figure}[b]
\begin{center}
\mbox{
\epsfysize=7.65cm
\epsffile{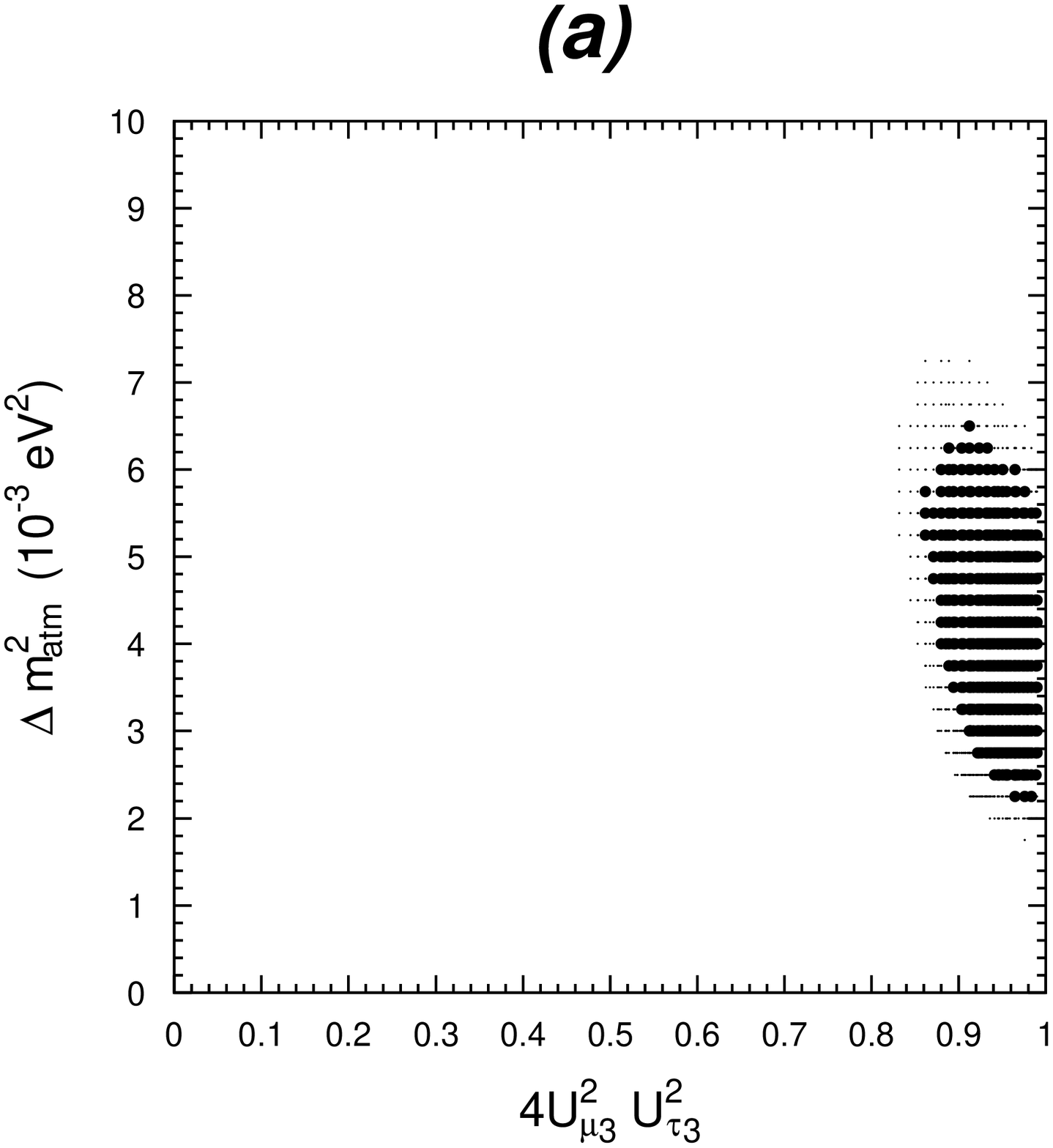}
}
\mbox{
\epsfysize=7.65cm
\epsffile{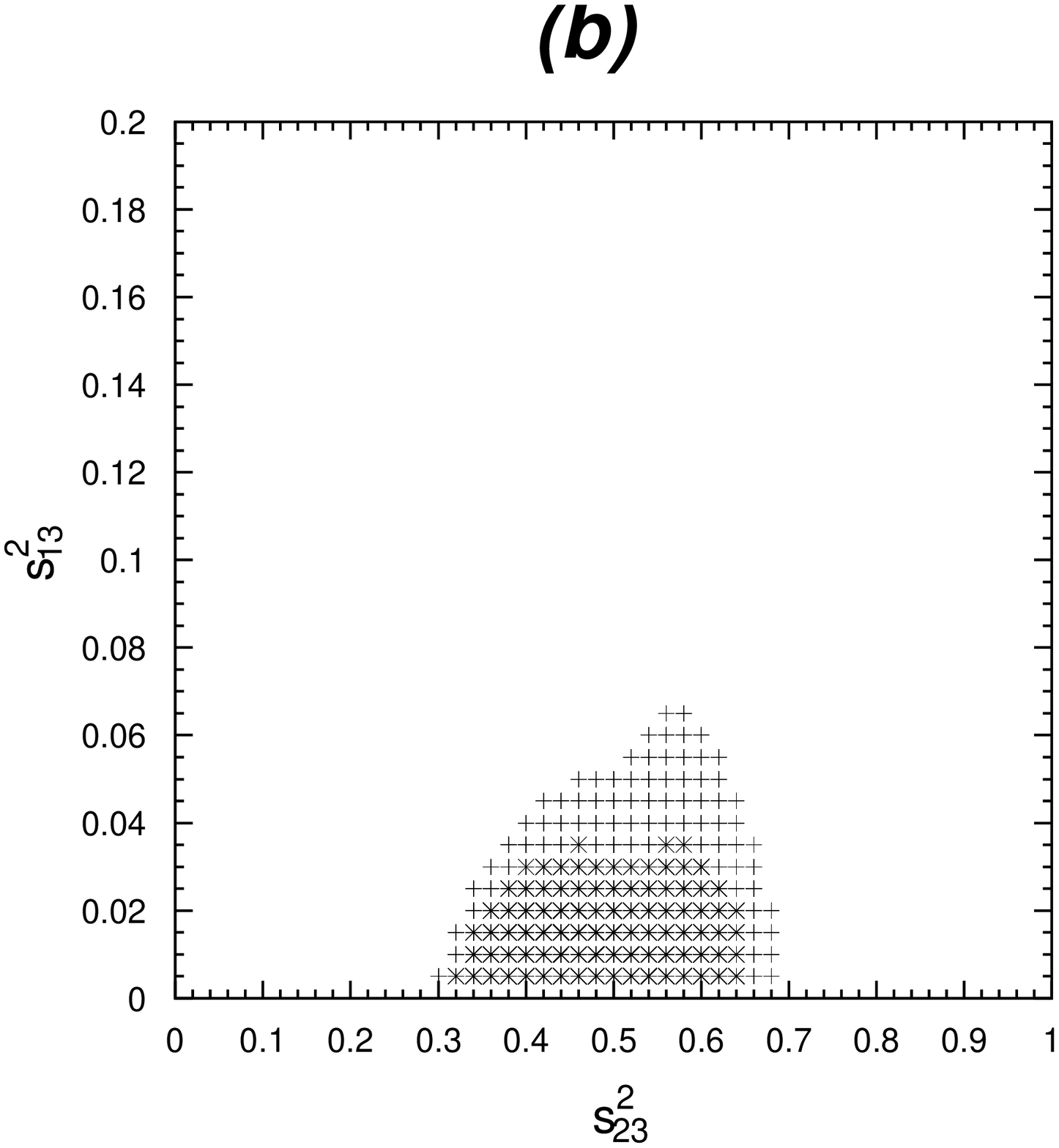}
}
\end{center}
\vspace{-0.5cm}
\caption{The allowed region in the conventional case. (a) for the parameters 
$\D m^2_{\rm atm}$ and 
$4U_{\mu 3}^2U_{\tau 3}^2 \simeq \sin^2(2\theta_{\rm atm})$ 
with 90\% (99\%) CL marked with a $\bullet$ ($\cdot$); 
(b) for the parameters 
$s_{13}^2$ and $s_{23}^2$, with 90\% (99\%) CL marked by a + 
($\stackrel{}{*}$). 
We have used CHOOZ and SK atmospheric data and fitted with 
four parameters. 
} 
\label{con2}
\end{figure}

Figure \ref{con2}(a) shows the allowed region in 
$\D m^2_{31}=\D m^2_{\rm atm}$ and 
$4 U_{\mu 3}^2U_{\tau 3}^2 \approx \sin^2(2\theta_{\rm atm})$. 
The confidence  intervals are calculated for 
four parameters (as we keep $s_{12}$ fixed),
so that the 90\% (99\%) confidence interval corresponds to 
$\chi^2-\chi^2_{\rm min}<7.8\; (13.3)$ \cite{pdg}. 
At 99\% CL we find 
$1.75 \times 10^{-3}{\rm eV}^2<\D m^2 <7.5\times 10^{-3}$ eV$^2$
and $\sin^2(2\theta_{\rm atm})>0.83$. These bounds are in good 
agreement with previous analyses \cite{threefit,gonz}.

In figure \ref{con2}(b) we show the 99\% CL 
and the 90\% CL  
area in $s_{13}^2$ and $s_{23}^2$. 
At 99\% CL the upper bound on $s_{13}^2$ is 0.065, 
corresponding to a value of 
$\sin^2(2\theta_{\rm CHOOZ}) \equiv 4 U_{e3}^2 ( 1 - U_{e3}^2)$ 
smaller than 0.25.

\begin{figure}[t]
\begin{center}
\mbox{
\epsfysize=7.65cm
\epsffile{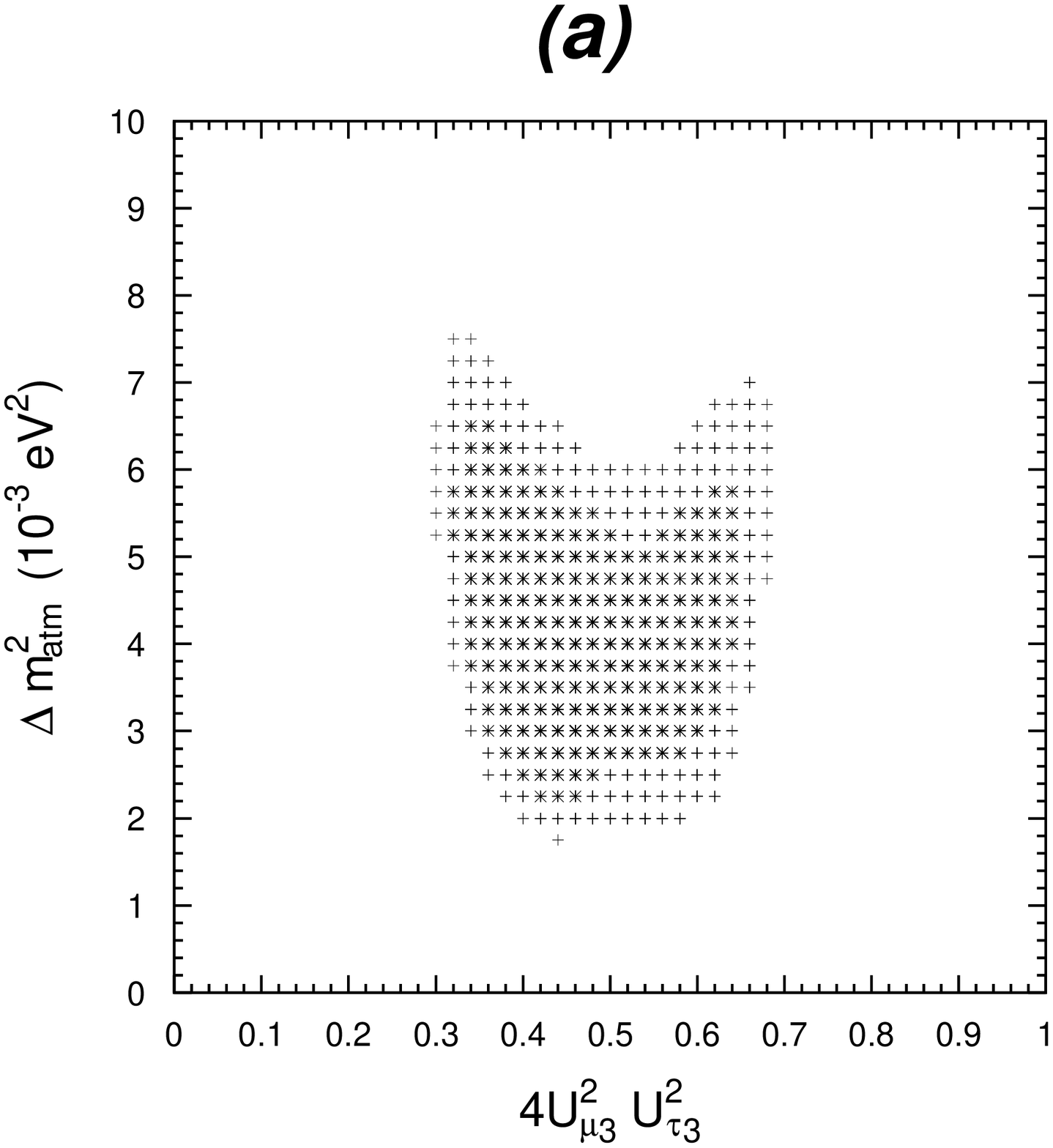}
}
\mbox{
\epsfysize=7.65cm
\epsffile{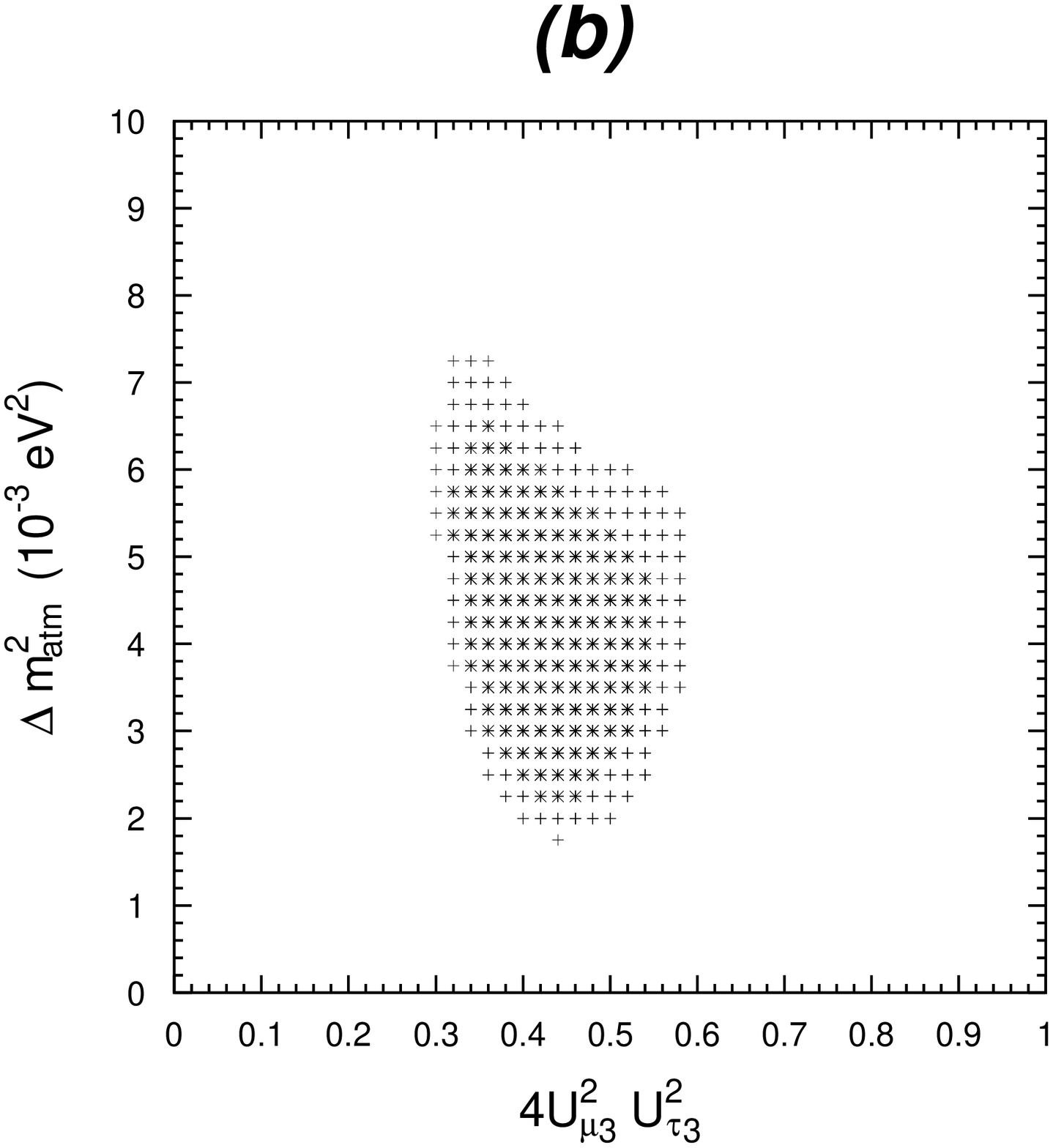}
}
\end{center}
\caption{Allowed region in $\D m^2_{\rm atm}$ and $s_{23}^2$ for 
the conventional 
case. In both figures the + ($+ \!\!\!\!\!\!\:{\times}$) 
indicates points within 99\% (90\%) CL.
(a) for all other parameters unconstrained; 
(b) with $\D m_{21}^2=2\times 10^{-4}$. 
We have used CHOOZ and SK atmospheric data and fitted with 
four parameters. } 
\label{con1}
\end{figure}

The full three-neutrino fit is not symmetric around 
$s_{23}^2=0.5$, as is also seen in Fig.~\ref{con1}(a) and (b). 
The small asymmetries arise from non-zero values of 
either $s_{13}$ or $\dmsq_\odot$.
This may be interpreted in terms of the electron excess
(see the discussion in Sec.~\ref{more-atm}).

We now consider the combined fit including the effect of the solar neutrino 
data in the LMA region. 
The common parameter, $\dmsq _\odot$, 
must be taken to be the same, and we calculate $\chi^2_I$  for a few values 
of this parameter. The minimum value occurs at 
$\dmsq_{\odot}=8 \times 10^{-5}$ with 
$\chi^2_{\rm LMA}=37.3$ \cite{gonz}. Hence when 
fitting to all three experiments 
(SK atmospheric, CHOOZ, solar data), 
we get $\chi^2_{I({\rm min})} \approx 84$ with 6 
parameters and 97 data points, 
so that $\chi^2_{I({\rm min})}/{\rm dof} \approx$ 0.93.

Recently a fit was performed to the solar data allowing for a free B 
and {\it hep} flux in the Sun \cite{newsolar}. The obtained LMA $\chi^2$ 
was 29.0 for 39 data points with 4 fit parameters, $\dmsq_\odot,
\theta_\odot$ and the fluxes of $^8$B and $hep$ neutrinos. 
Combining this with our fit for SK and CHOOZ, 
we would get $\chi^2_{I({\rm min})} \approx$ 78 
, so that $\chi^2_{I({\rm min})}/{\rm dof} \approx$ 
0.90.  
Allowing for an overall normalization of the atmospheric fluxes, as done 
by the SK collaboration, could in general also improve the goodness of 
fit for atmospheric data. 
Let us also note that in Ref. \cite{gonz} a combined fit to 
atmospheric, solar and CHOOZ data gave a best fit 
of\footnote{The difference from our fit is that in \cite{gonz},
the $\dmsq_{\odot}$ effects on $\chi^2_{\rm atm}$ are neglected, 
whereas we neglect $s_{13} \neq 0$ effects on $\chi^2_{LMA}$.}
$\chi^2/{\rm dof}=0.97$.

The LMA solution currently gives the best fit to the solar data, and the 
fit to the atmospheric data is not affected much by very small 
solar mass differences. Therefore we expect that the value of $\chi^2$ can 
only be higher in other regions of parameter space than the one 
computed above.

\subsection{Fit to atmospheric, LSND and CHOOZ data (scheme II)}
\label{atm-lsnd}

In this section we give the results of fitting 
to the SK atmospheric, CHOOZ and LSND data and hence will take
$\D m^2_{21} = \D m^2_{\rm atm}$ and 
$\D m^2_{31} \simeq \D m^2_{32}=\D m^2_{\rm LSND}$. 
The SK and CHOOZ data are largely 
independent of the large $\D m^2$, which we can therefore freely use 
to fit the LSND experiment.  
For every point in parameter space we simply choose to calculate the 
value of $\D m_{31}^2$ that fits the LSND data best. 
The remaining four parameters ($s_{12}, s_{23}, s_{13}, \dmsq_{21}$) 
are varied to obtain confidence levels.

\begin{figure}[t,b,h]
\begin{center}
\mbox{
\epsfysize=7.65cm
\epsffile{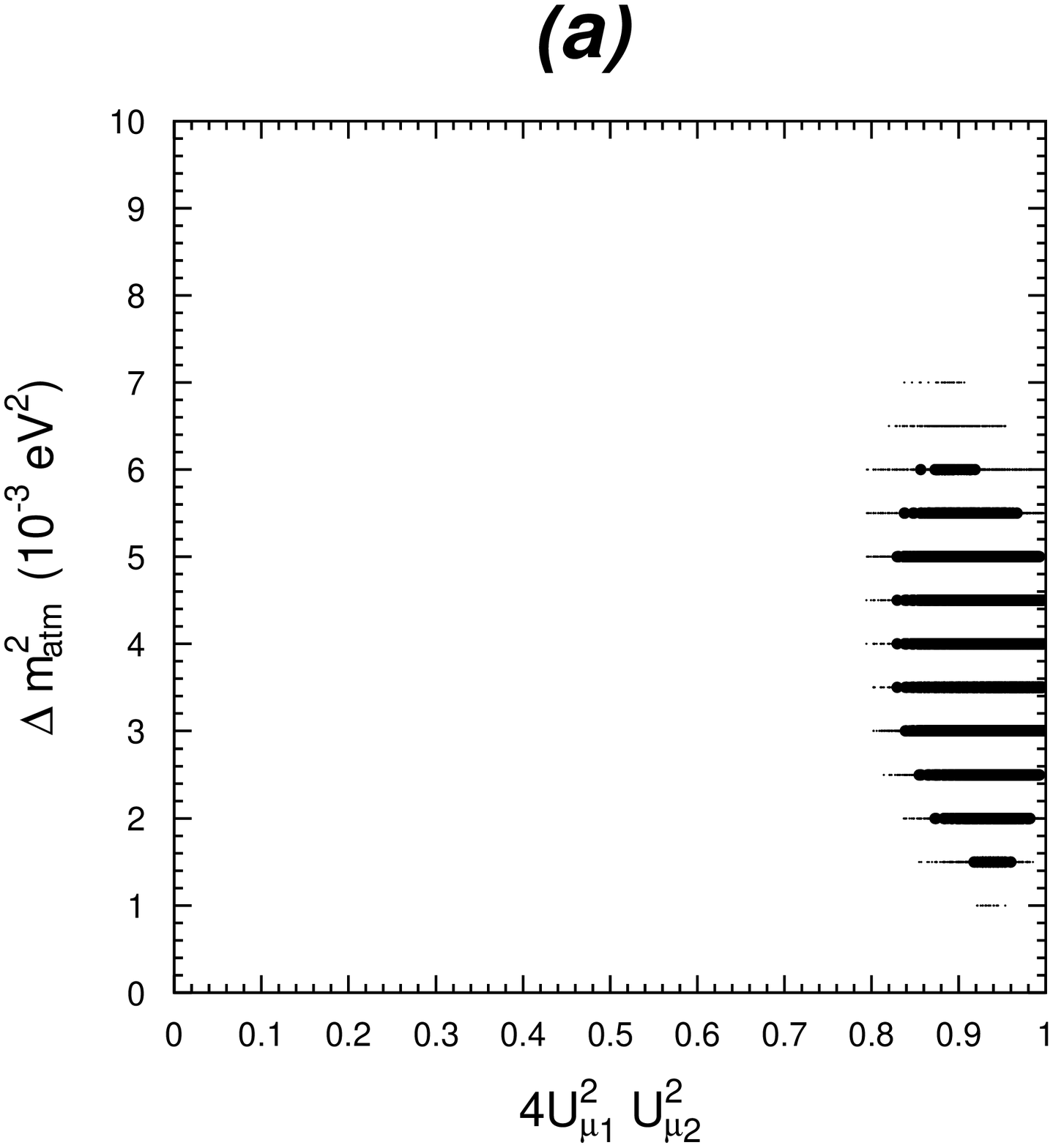}
}
\mbox{
\epsfysize=7.65cm
\epsffile{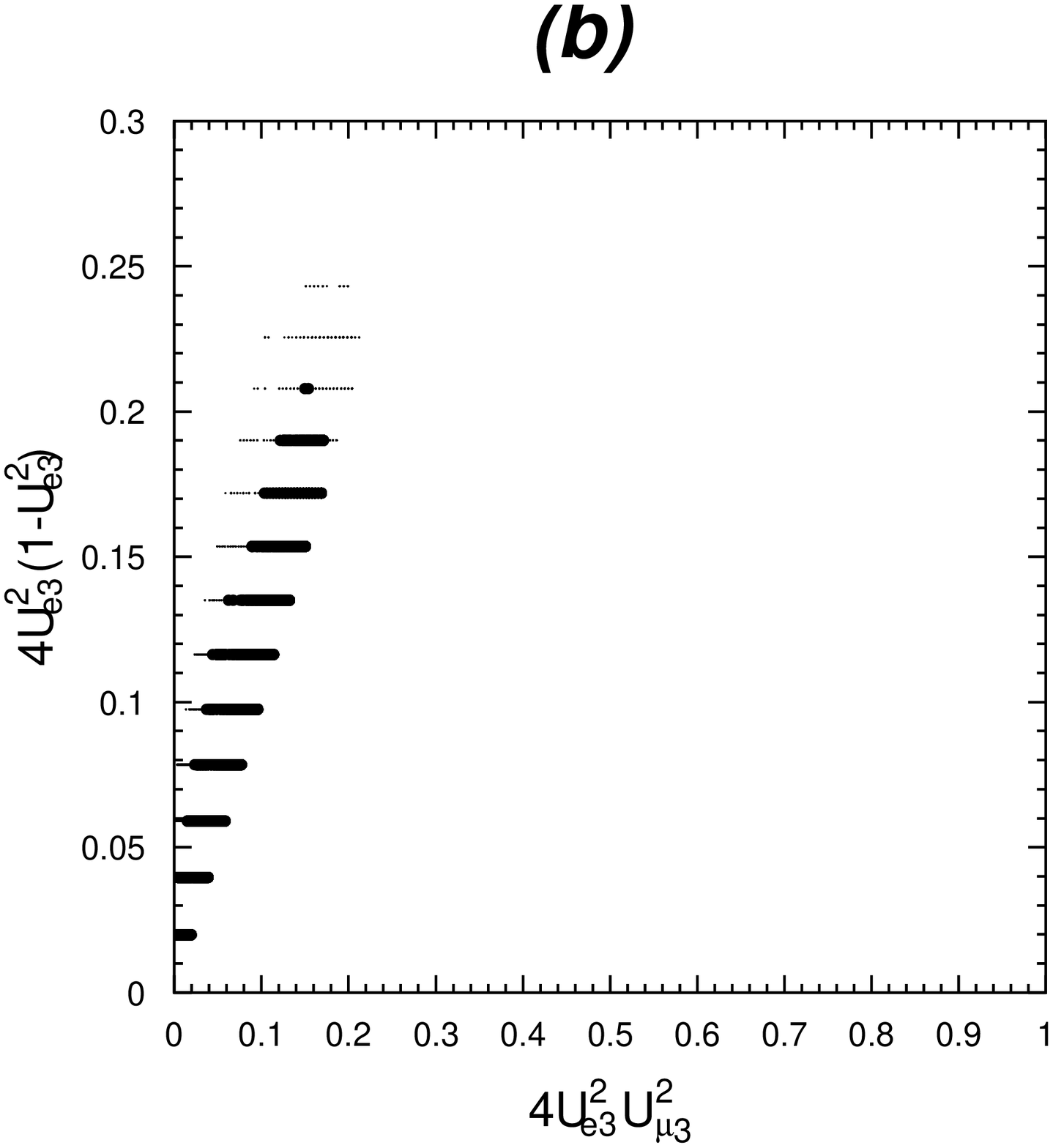}
}
\end{center}
\vspace{-0.5cm}
\caption{Allowed region for scheme II in (a) $\dmsq_{\rm atm}$ and 
$4U_{\mu 1}^2 U_{\mu 2}^2 \simeq \sin^2(2\theta_{\rm atm})$
and (b) $4U_{e3}^2(1-U_{e3})^2 \simeq \sin^2(2\theta_{\rm CHOOZ})$ and 
$4U_{e3}^2 U_{\mu 3}^2 \simeq \sin^2(2\theta_{LSND})$. 
Points within 90\% (99\%) CL are marked by a $\bullet$
($\large{\cdot}$). We have used CHOOZ and SK atmospheric data and 
fitted with four parameters. 
} 
\label{lsnd1}
\end{figure}

Let us first perform a combined fit to the SK and CHOOZ data with
$\chi^2 \equiv \chi^2_{\rm atm} + \chi^2_{\rm CHOOZ}$, as we
did for scheme I in Sec.~\ref{atm-sol} (again we have 55 data points).
The confidence levels are calculated for four fitting 
parameters. 
Fig.\ref{lsnd1}(a) shows the allowed region in $\D m_{\rm atm}^2$ and 
$4U_{\mu1}^2 U_{\mu 2}^2\simeq \sin^2(2\theta_{\rm atm})$. 
At 99\% confidence level 
the bounds are $4U_{\mu1}^2 U_{\mu 2}^2 > 0.8$
and 1 eV$^2 < \D m^2_{\rm atm} < 7$ eV$^2$. The allowed region is very 
similar to the one found in the conventional case. It shows that the 
two-neutrino atmospheric neutrino fit is only slightly affected 
by the three-neutrino extension in both schemes. 
Although we note that the region in $\D m^2_{\rm atm}$ 
is lowered in scheme II. This is due to the small 
constant contributions that arise from the averaging of 
the $\sin^2(\D m^2_{\rm LSND} L/4E)$ terms.

Let us now include the LSND data in the fit.
The best-fit point with $\chi^2_{II({\rm min})}=44.9$ for the combined fit 
is obtained at
\be\label{lsndbf} 
 s_{12}^2=0.69 \;,\; s_{23}^2=0.98 \;,\;s_{13}^2=0.98 \;,\; 
\Delta m_{21}^2=0.0031  \;, \;\Delta m_{31}^2=0.22 \;.  
\ee
This should be compared with the two-neutrino $\nu_\mu \rightarrow \nu_\tau$ 
minimum value of $\chi^2$ of 49.5 (at $\dmsq_{\rm atm}=0.0038$). 
We remind the reader that the LSND datum is chosen to be fitted best. 
Hence the effects of the three-neutrino scheme are not large, but still 
significant. The main reason for the lowering of the $\chi^2$ is an 
excess in both sub-GeV and multi-GeV 
electron neutrino ratios. The goodness of fit to the data, having 
$\chi^2_{\rm II}/{\rm dof} \approx 0.88$ 
is thus as good as the one obtained to
the solar, atmospheric and CHOOZ data in Sec.~\ref{atm-sol}.

The LSND probability is
\be
 P_{\bar{\nu}_\mu \rightarrow \bar{\nu}_e} \simeq  
4 U_{e3}^2 U_{\mu 3}^2 \sin^2 \left( \frac{\D m_{\rm LSND}^2L}{4E} \right) 
   -4U_{e1}U_{e2}U_{\mu 1}U_{\mu 2} \sin^2
\left( \frac{\D m_{\rm atm}^2L}{4E} \right) \;,
\label{bf-II-1}
\ee
where the second term is small because of the small value of 
$\sin^2(\dmsq_{\rm atm}L/4E)$, so that we may define the LSND angle 
as $\sin^2(2 \theta_{\rm LSND}) \equiv 4 U_{e3}^2 U_{\mu 3}^2$.
In terms of physical angles, the best-fit point corresponds to 
$\sin^2(2\theta_{\rm atm}) = 0.99$ and 
$\sin^2(2 \theta_{\rm LSND}) \approx 0.08$.

The best-fit point (\ref{bf-II-1}) has an LSND angle in the region 
excluded by Bugey \cite{bugey}. 
We investigate the impact of the Bugey data on the 
value of the LSND angle when combining these in the fit. 
Including the Bugey data gives a best-fit point 
with $\chi^2_{II({\rm min})}= 47.0$ at
\be 
s_{12}^2=0.7 \;,\; s_{23}^2=0.98 \;,\; s_{13}^2=0.985 \;,\; 
\dmsq_{21}=0.0035 \;,\; \dmsq_{31}=0.26 ~.
\label{bf-II-2}
\ee
or in terms of physical angles 
$\sin^2(2\theta_{\rm LSND})\simeq 0.06,
\; \sin^2(2\theta_{\rm atm}) \simeq 0.98$.
With 59 data points fitted with 5 parameters, we have 
$\chi^2/{\rm dof} \approx$ 0.87. 
The upper bound at 90\% CL is $\sin^2(2\theta_{\rm LSND})<0.11$. 
Therefore the inclusion of the Bugey data 
still favours large LSND angles.

Also the best-fit point (\ref{lsndbf}) is slightly different from the 
one obtained in \cite{choubey}; in particular the LSND angle is larger. 
It should be said that in Ref. \cite{choubey} the LSND data were not 
included in the fit, but only a mass squared difference of order eV$^2$.  
We therefore believe that the discrepancy is due to the 
restriction $\dmsq_{LSND}>0.5$ in \cite{choubey}, which is equivalent 
to restricting the value of the LSND angle to 
$\sin^2(2\theta_{\rm LSND})<0.02$, and also to the use 
of up/down asymmetries as representing atmospheric data.

Since $U_{e3} = s_{13} \approx 1$ in scheme II, the solutions 
are still close to an effective two-neutrino mixing between
$\nu_\mu$ and $\nu_\tau$, with $\nu_e$ dominating the third
mass eigenstate. For the normal hierarchy (discussed here),
$\nu_e$ is almost equal to the heaviest state, whereas for the
inverted hierarchy, $\nu_e$ is almost equal to the lightest state.

Another interesting point about scheme II is the connection 
between the ``LSND angle'' and the ``CHOOZ angle''.
The LSND angle has already been defined above.
In order to define the ``CHOOZ angle'', let us have a look at 
the CHOOZ survival probability for $\nu_e$. 
Since the matter effects are small, for the analytical 
approximations we may put the parameters equal to their 
vacuum values. 
The $\nu_e$ survival probability can be approximated by 
\bea
     P_{\nu_e \rightarrow \nu_e} & = & 1-4U_{e3}^2 (1-U_{e3}^2) 
      \sin^2 \left( \frac{\Delta m_{LSND}^2L}{4E} \right) 
      -4 U_{e1}^2 U_{e2}^2 \sin^2 
\left( \frac{\Delta m_{\rm atm}^2 L}{4E} \right) 
	\nonumber \\ 
       & \simeq & 1-2U_{e3}^2(1-U_{e3}^2) \quad .
\eea
The dominant term in the survival probability $P_{ee}$
originates from the $\dmsq_{\rm LSND}$ term,
since the coefficient in front of the $\dmsq_{\rm atm}$ is very small
(of order $c_{13}^4$). 
Hence the CHOOZ angle is approximated by 
$\sin^2 (2 \theta_{\rm CHOOZ}) \approx 4U_{e3}^2(1-U_{e3}^2)$. 

Figure \ref{lsnd1}(b) shows the allowed 
region in the parameter space of
$\sin^2(2\theta_{\rm CHOOZ})$ and $\sin^2(2\theta_{\rm LSND})$. 
The lower (upper) bound on the LSND angle then turns into a 
lower (upper) bound on the CHOOZ angle. 
When considering the combined fit to the atmospheric, CHOOZ 
and LSND data,   
the upper bound to 90\% CL with 5 parameters is
$\sin^2(2\theta_{\rm CHOOZ})< 0.20$.
The corresponding lower bound is
$\sin^2(2\theta_{\rm CHOOZ})> 1.2 \times 10^{-3}$.
Although the bound is very weak, it does require in principle 
a full three-neutrino mixing scheme, and will be important 
in future experiments trying to extract CP-violation.

Let us briefly note that the scheme suggested in \cite{sb}, which 
attempts to solve all three neutrino anomalies with just three flavours,  
is found to be strongly disfavoured, in agreement with previous results 
\cite{sbnot}. Within this scheme, $\chi^2$ is always above 120. Recently 
the authors of \cite{gupta} also suggested two different regions, which 
they found to be good candidates for solving all three anomalies.   
We find these regions strongly disfavoured; in particular 
the region suggested in Table 2 of \cite{gupta} conflicts 
with the CHOOZ data and can be considered ruled out. 
The $\chi^2$ of the SLMA solution defined in \cite{schlatt} is also 
too large to be consistent with the data: 
at $s_{23}^2 = 0.5$, which is the best-fit point in this scheme,
$\chi^2_{\rm SLMA} = 166$.

To summarize, a good fit ($\chi^2/{\rm dof} \approx 1$) may be
obtained for the atmospheric, LSND and CHOOZ data
through scheme II. We would like to stress that the best-fit region 
has large LSND angles $\sin^2(2\theta_{\rm LSND}) \simeq 0.05$.  
Therefore the LSND 
mass squared difference might have observable effects in future 
measurements, as we would like to discuss next.

\section{Distinguishing between the schemes}
\label{compare}

Although comparing the $\chi^2$'s in the two schemes in order to
determine which is the better one is not valid,
we note that the individual $\chi^2$'s per degree of freedom are
near 1 and hence the data seem to be fitted reasonably well.
That these $\chi^2$/dof are similar indicates that
the goodness of fit for the two schemes are similar, and
by itself provides no grounds for preferring one scheme over the other.
The reason why most authors choose 
to disregard the LSND results is of course because it still needs 
confirmation. 
Since the current data do not seem to prefer either of the two
schemes, let us examine whether the future data will help to shed some 
light on the identification of the actual mass pattern.

\subsection{Electron neutrino excess in the atmospheric data}
\label{more-atm}

The muon neutrino deficit observed in the atmospheric neutrino
experiments can be explained well by both the schemes that we
are considering. In fact the amount of deficit observed 
is the dominant factor in the determination of 
the parameters in the fit. Although we might note that, in the region 
favoured in scheme II, the LSND angle also contributes to a lowering of the 
down-going muon neutrinos in the multi-GeV range. 
In fact small deficits are indicated by the data and because of the 
small stastistical errors on the multi-GeV muon neutrino ratios the effect 
contributes to the preference for a large LSND angle. Let us now examine the 
predictions of this scheme about a $\nu_e$ excess or deficit. The data show a 
small excess, which is accounted for by the SK collaboration by an overall 
flux normalization. The electron excess may be defined as
\be
\xi_e \equiv \frac{N_e}{N_e^0} - 1 = 
P_{ee} +  r P_{e \mu} -1 ~,
\ee
where $N_e \;(N_e^0)$ is the number of electron neutrinos
in the presence (absence) of oscillations, and $r$ is the ratio of
the original muon and electron neutrino fluxes. The value of $r$
depends on the energy as well as on the zenith angle.
For the sub-GeV neutrinos, $r \approx 2$ independent of 
the zenith angle, whereas for multi-GeV neutrinos, 
$r \approx 2$ for the near-horizontal direction and 
$r \approx 3$ for the vertical (either 
up- or down-going) neutrinos. 

In scheme I, there are two factors that determine the
extent of $\nu_e$ excess: $\D m^2_\odot$ and $s_{13}$.
The excess due to a non-zero value of $\D m^2_\odot$ may be
written in the form \cite{dmsq}
\be
\xi_{e(1)} = P_{\dmsq} (r c_{23}^2 - 1) \;,
\label{xie1}
\ee
whereas the excess due to non-zero $s_{13}$ may be approximated
as \cite{amol}
\be
\xi_{e(2)} = P_{s_{13}} (r s_{23}^2 - 1) \;.
\label{xie2}
\ee
The value of $\theta_{23}$ affects $\xi_{e(1)}$ and $\xi_{e(2)}$ 
in opposite directions: with an increasing value of $\theta_{23}$, 
the excess due to the first (second) term decreases (increases).
Let us look at the two terms separately.

(i) If $\theta_{13} \approx 0$, the solar mass squared difference  
will produce an excess (deficit)
for low (high) values of $s_{23}$. For the sub-GeV data 
this crossing occurs at  $s_{23}^2=0.5$. 
Furthermore the excess has the following energy 
and zenith angle dependence \cite{dmsq}: 
For the sub-GeV ratios the excess can be as large as 10-12\% 
for large values of $\dmsq_{\odot}$ with a positive up/down asymmetry but  
weakly depending on the zenith angle. For the multi-GeV ratios the excess is 
small (5-7 times smaller than for sub-GeV) and the up/down asymmetry 
is positive. 
The excess is negligible for $\dmsq_{\odot}< 4 \times 10^{-5}$.

(ii) If $\dmsq_{\odot} \approx 0$, then a non-zero value of $s_{13}$ 
will result in an excess (deficit) for high (low) values of $s_{23}$. 
For multi-GeV up- and down-going neutrinos, this zero crossing point 
occurs at $s_{23}^2 \approx 0.3$.
The energy and zenith angle dependence can be 
described as \cite{amol}: 
Only in the multi-GeV range and for up-going neutrinos can the excess 
be substantial and hence the up/down asymmetry is positive. 
    
\begin{figure}[t,b,h]
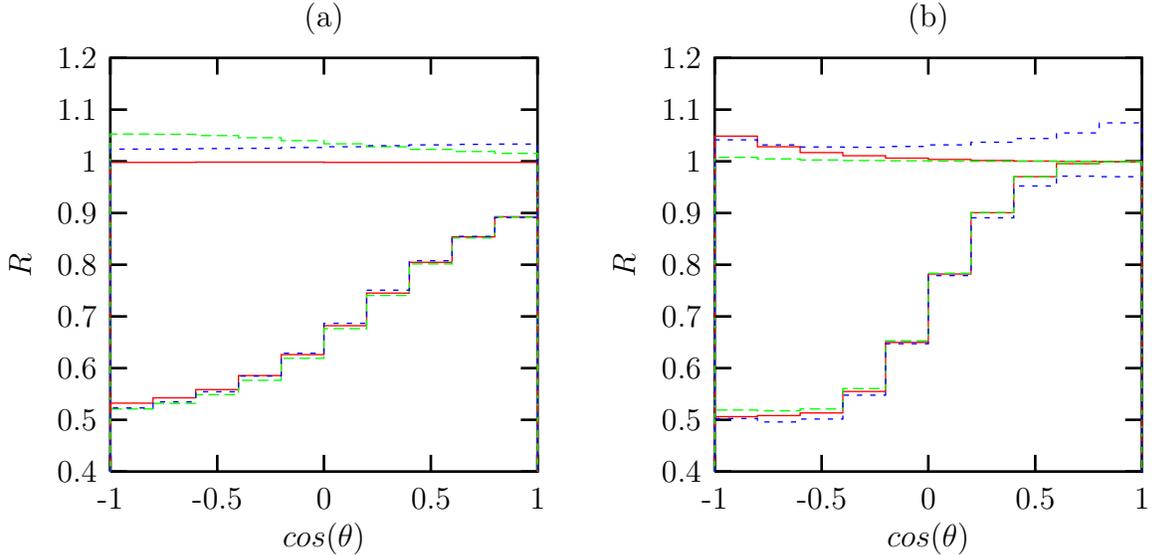

\mbox{
\input{ratio_s.ps}
}
\mbox{
\input{ratio_m.ps}
}
\caption{Ratios as a function of the zenith angle for three best-fit points 
for (a) sub-GeV and (b) multi-GeV. 
The long-dashed curve is for scheme I constrained to 
$\dmsq_{\odot}=2\times 10^{-4}$. The solid curve is for scheme I 
constrained to $\dmsq_{\odot}=2 \times 10^{-5}$. The short-dashed curve 
is for scheme II. 
Upper curves are electron neutrino ratios and lower curves are muon neutrino 
ratios.
} 
\label{ratiofig}
\end{figure}

For both effects we therefore have that for down-going neutrinos in the 
multi-GeV range the electron excess vanishes. Also an excess is 
accompanied with 
a positive up/down asymmetry. Deficits, though, could give negative 
up/down asymmetries, but this is disfavoured by the data. 
Note that for high values of $\dmsq_\odot$ the best-fit point 
were found to be near $s_{13}\simeq 0$ and $s_{23}\simeq 0.4$, 
whereas for small $\dmsq_\odot$  the best fit was 
for non-zero (albeit small)
values of $s_{13}$ and $s_{23}\simeq 0.6$ (Sec.~\ref{atm-sol}). 
The data therefore seem to prefer a small excess in both the 
sub-GeV and multi-GeV ranges. 
The ratios for these best-fit points are plotted in Fig.\ref{ratiofig}.

\begin{figure}[t,b,h]
\begin{center}
\mbox{
\epsfysize=7.65cm
\epsffile{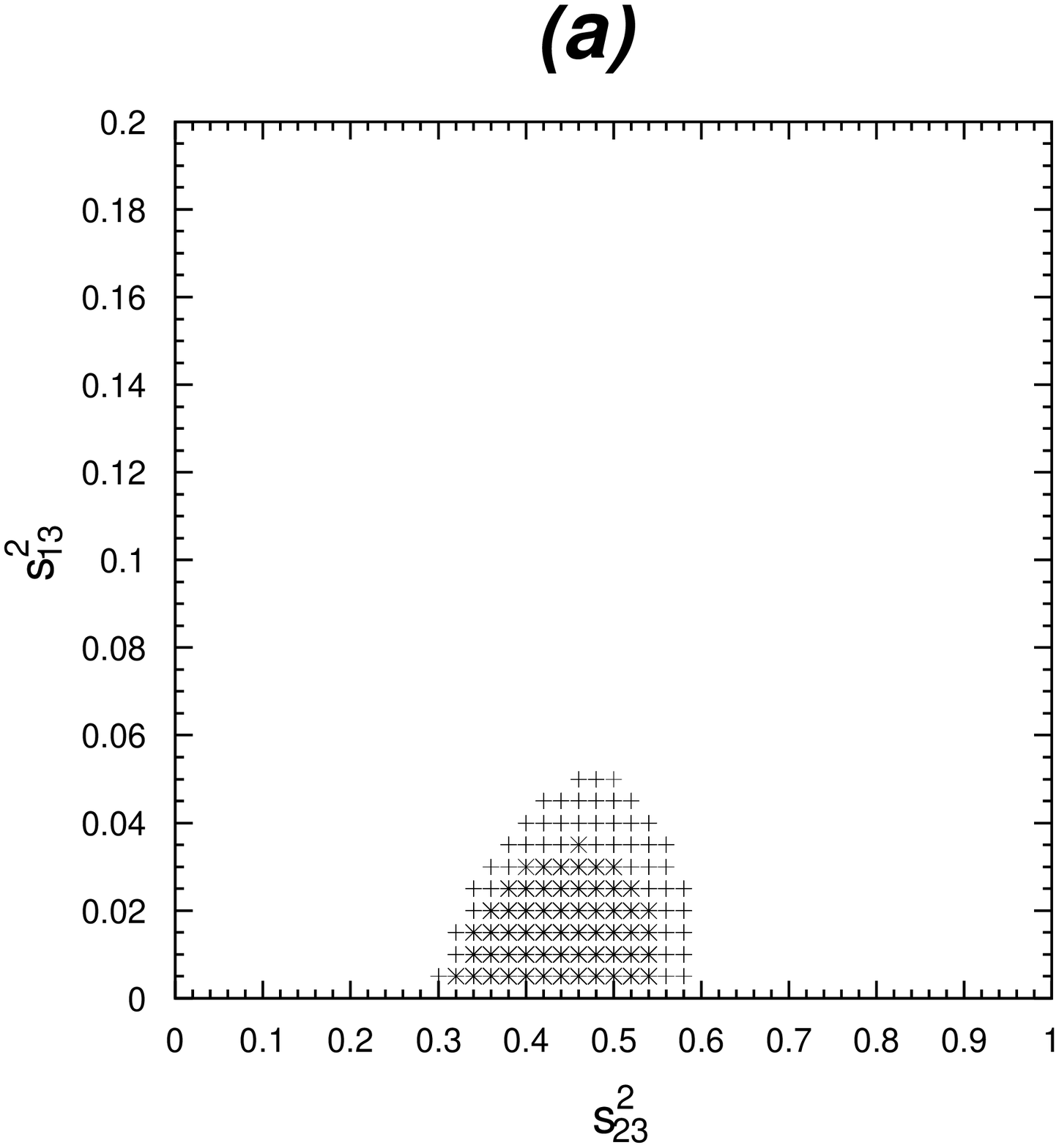}
}
\mbox{
\epsfysize=7.65cm
\epsffile{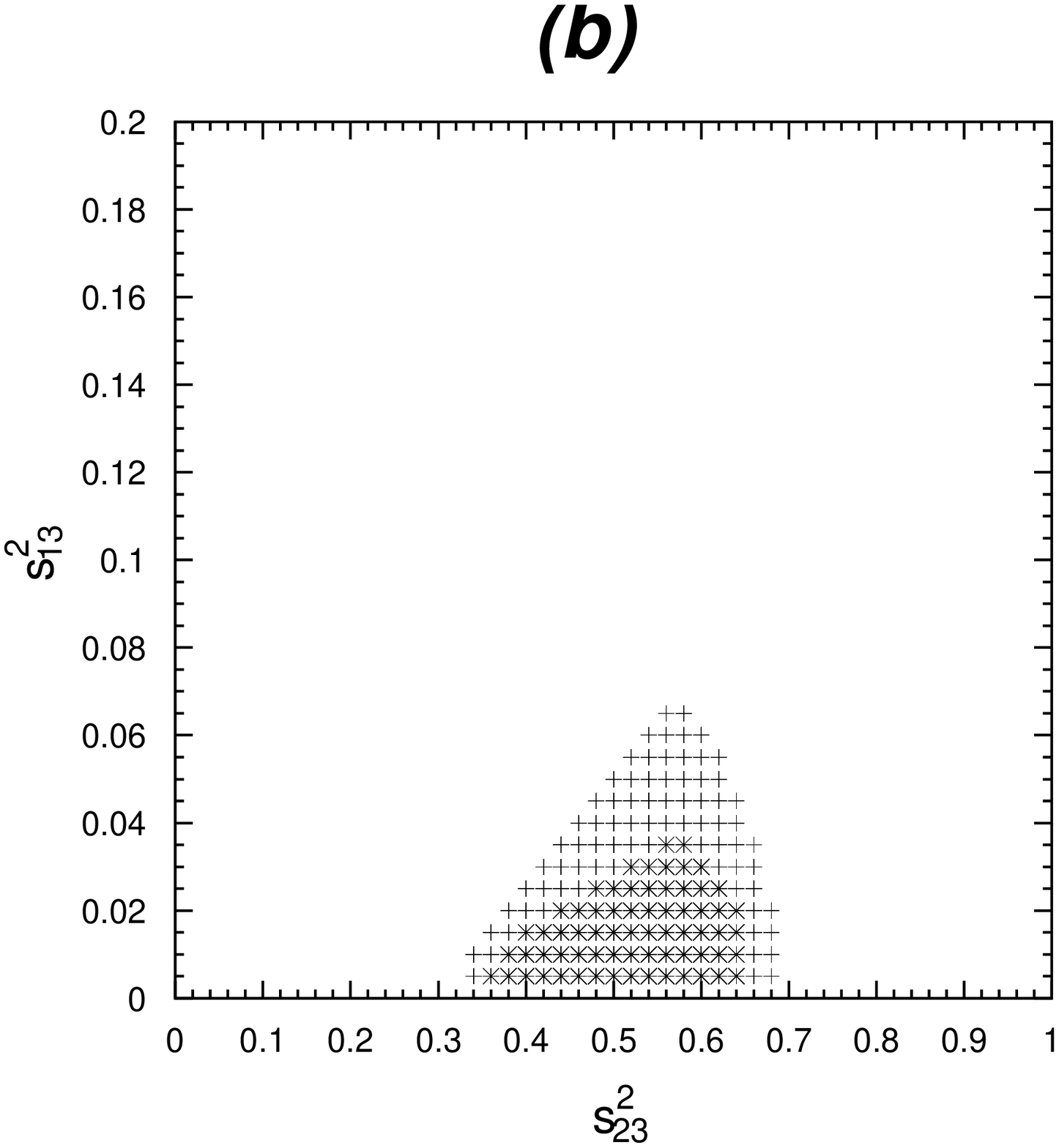}
}
\end{center}
\vspace{-0.1cm}
\caption{ The allowed region in $s_{13}^2$ and $s_{23}^2$ for scheme I for 
$\dmsq_{\odot}$ constrained to    
(a)~$2 \times 10^{-4}$ and (b) $2 \times 10^{-5}$. 
We have used CHOOZ and SK atmospheric data and fitted with four parameters. 
} 
\label{s13fig}
\end{figure}

Large values of $\dmsq_\odot$ would tend to produce a deficit at high
$s_{23}$ for sub-GeV neutrinos. 
Since the data actually show a slight excess, high values 
of $s_{23}$ are not favoured with large $\dmsq_\odot$. This may
be noted from Fig.~\ref{s13fig}(a), which shows that 
the SK atmospheric and CHOOZ data disallow nearly all the 
``dark side'' ($s_{23}^2 > 0.5$) for $\dmsq_{\odot}=2\times 10^{-4}$.
On the other hand, Fig.~\ref{s13fig}(b) shows that the observed excess 
goes on to allow larger values of $s_{13}$ for small solar mass differences 
on the dark side, where the produced excess is positive. 
The present data are however insufficient for us to be able to detect 
any sign of a possible non-zero value of either $s_{13}$ or $\dmsq_{\odot}$. 

\begin{figure}[h]
\begin{center}
\mbox{
\epsfysize=9cm
\epsffile{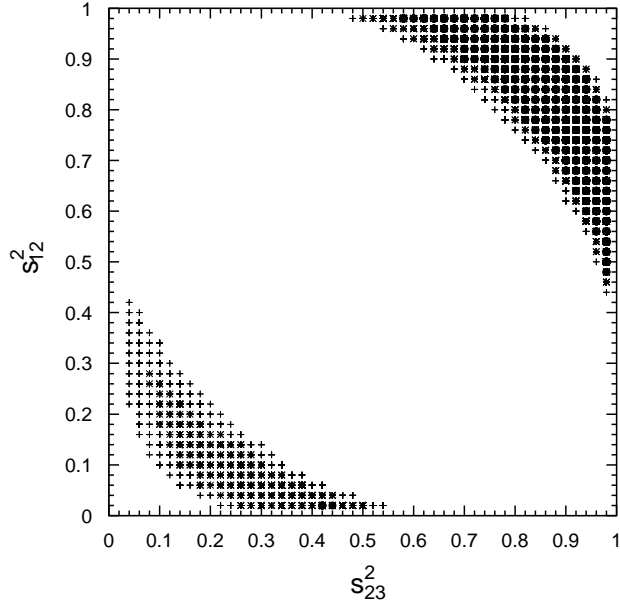}
}
\end{center}
\vspace{-1cm}
\caption{Allowed region in scheme II for 
parameters $s_{12}^2$ and $s_{23}^2$.  
Points within 68\% (90\%, 99\%) CL are marked by 
$\bullet$ ($+ \!\!\!\!\!\!\:{\times}$,+). 
We have included the LSND, CHOOZ and SK atmospheric data and fitted with 
four parameters.}
\label{lsnd2}
\end{figure}

In scheme II, the oscillations due to $\dmsq_{LSND}$ are
averaged out and the electron excess becomes
\be
\xi_e \approx  
2 s_{13}^2 c_{13}^2 (r s_{23}^2 -1) 
+  2 r c_{13}^2 s_{12} c_{12} \sin(2\theta_{12}+ 2\theta_{23})
\sin^2 \left( \frac{\D m_{\rm atm}^2L}{4E} \right)~~,
\label{xi2}
\ee
there are no matter effects and the angles and $\dmsq$'s are the same
as their vacuum values. 
We have neglected the terms that are more than second order
in the small quantity $c_{13}$. For the upper right corner of 
Fig.\ref{lsnd2}, we have 
$\sin(2\theta_{12}+ 2\theta_{23}) \simeq \sin(2\theta_{\rm atm}) \simeq -1$ 
and for the lower left corner we have 
$\sin(2\theta_{12}+ 2\theta_{23}) \simeq 1$, 
so the sign of the second term in (\ref{xi2}) can change.
Also, the sign of the first term can change, depending on the 
value of $s_{23}$ and 
$r$, determined by energy range and zenith angle. Small values of 
$s_{23}$ are correlated with small values of the LSND angle, 
although the exact value also depends on $s_{13}$  
($s_{23}^2<0.4$ is disallowed as seen in Fig. \ref{lsnd2} 
since the LSND angle becomes too small).     
The best-fit points found in Sec.~\ref{atm-lsnd} had 
$s_{23} \approx 1$, and the values with $s_{23}^2 > 0.5$
are favoured in the fit.

Assuming we are in the area with $s_{23}^2>0.5$, 
the two terms thus contribute to the electron 
ratio in opposite directions.
The first term in (\ref{xi2}) is identical for up- and down-going 
neutrinos, when neglecting the small asymmetry caused by the 
magnetic field of the Earth, and produces an excess. 
The second term vanishes for down-going neutrinos
due to small $L$, whereas it gives a small negative contribution 
for up-going neutrinos. The up/down asymmetry is
then negative. Furthermore the magnitude of the excess is larger 
for multi-GeV neutrinos, since the value of $r$ is larger.
The magnitude depends on the interplay between the LSND and the CHOOZ 
angles. The maximum for a given CHOOZ angle is
$(r-1)\sin^2(2\theta_{\rm CHOOZ})/2$. 
For $\sin^2(2\theta_{\rm CHOOZ})=0.08$ the maximum excess is around 4\% 
in the sub-GeV range and around 6\% in the multi-GeV range.   
For $\sin^2(2\theta_{\rm CHOOZ})=0.03$ the effect is only substantial in the 
multi-GeV range and is around 2\%.
The sign of the up/down asymmetry is opposite to
that obtained in scheme I for both sub-GeV 
and multi-GeV (disregarding the possibility of a negative asymmetry 
along with a deficit in scheme I ). 
It should be noted that for small values of the LSND angle a 
small positive up/down asymmetry, in both sub-GeV and multi-GeV ranges, 
can be obtained in scheme II, again normally accompanied by deficits. 
Nevertheless for some points within the lower left corner having 
$s_{23}^2\simeq 0.5$, a positive up/down asymmetry is obtained along with 
an excess in both sub- and multi-GeV ranges. Hence unfortunately the 
predictions about the electron ratios are not unique is scheme II.

In the preferred region of scheme II the largest excess is expected 
to be for down-going multi-GeV neutrinos (see Fig.~\ref{ratiofig}), 
whereas in scheme I this ratio is one\footnote{
The new calculation of the atmospheric neutrinos fluxes \cite{flux} 
indicates that the fluxes are slightly smaller than used by SK and 
the discrepancy is larger for higher energies. This would imply a larger 
electron excess in the multi-GeV range and would therefore seem to prefer 
scheme II. It could also result, though, in an approach to one of 
the down-going muon neutrinos ratio in the multi-GeV range.}. 
Let us briefly mention that the 
large value of $r$ is the reason why the muon neutrino ratios are 
affected less than the electron neutrino ones, in the former the 
$P_{\nu_\mu \nu_e}$ probability is multiplied by $1/r$.  
A clear experimental signal for an LSND mass squared difference is therefore 
an excess of down-going multi-GeV electron neutrinos along with a small 
muon neutrino deficit. 
With the present statistics, however, it is not possible to distinguish 
between the two schemes. 
Nevertheless, with the distinct pattern described above, it will be 
made possible by future precise measurements and 
improvements of the atmospheric flux calculation.

\subsection{Long baseline experiments}
\label{long-baseline}

Let us look at the possibility of distinguishing between the two 
schemes in long baseline experiments, where the initial neutrino
spectrum and flux are better known and hence one may expect to have
a better handle on the mixing parameters.

The K2K experiment in Japan started to report their first results 
\cite{k2k}. The almost pure $\nu_\mu$ beam (98.2\% 
$\nu_\mu$, 1.3\% $\nu_e$ and 0.5 $\bar \nu_\mu$) travels 250 km 
from the KEK laboratory to the SK detector. The average neutrino energy 
is $1.3$ GeV. Hence the value of $\sin^2(\dmsq_{\rm atm}L/4E)$ is 
large and in scheme II it will dominate over the LSND mass squared terms, 
which are accompanied by small angles. Also, the effect of 
$\dmsq_{\rm solar}$ is negligible in scheme I. 
This already tells us that the predictions for the two schemes 
will be nearly the same. In the following we will neglect the small 
contamination of electron neutrinos and antimuon neutrinos. 
The expected muon neutrino spectrum at SK, $\Phi_{\nu_\mu}$, 
has been reported in \cite{k2k1} after $10^{20}$ protons on target (pot). 
From this spectrum we calculate the number of muon and electron neutrino 
events by performing the integral
\be\label{lbl}
 N_{\alpha}= \int dE_\nu \sigma_{\alpha}(E_\nu) 
P_{\a \mu}(E_\nu) \Phi_{\nu_\mu}(E_\nu) \;,
\ee
where the probabilities have been calculated using a constant matter 
density of 2.7 g/cm$^3$ in the Earth's crust. First we compute the 
total number of $\nu_e$ and $\nu_\mu$ events for all points 
within 99\% CL 
to see if it can reveal a difference. The maximum number of $\nu_e$ 
events are found to be 25 for scheme I and 45 for scheme II. 
However already for $\sin^2(2\theta_{\rm LSND})=0.03$ the maximum 
number is the same. 
Therefore in order to see the difference the energy spectrum must be measured 
with a high precision and furthermore the LSND angle must be large.
Figure ~\ref{k2kfig} shows the expected number of 
$\nu_e$ and $\nu_\mu$ events as a function of energy for 
different points within the two schemes, all within 90\% CL. 
The short-dashed curve is within scheme II, and chosen to make the number of 
$\nu_e$ events small. The long-dashed curve, also within scheme II, 
is chosen so as to make the number of $\nu_e$ events large 
(equivalently a small number of $\nu_\mu$ events). 
Also two points within the conventional case (scheme I) are shown, 
again one giving a large number of $\nu_e$ events (dotted curve), 
and one with a small number of $\nu_e$ events (dot-dashed). 
The number of events in the conventional case falls rapidly above 
1 GeV, which is not always the case in the LSND scheme. Hence a signal for 
scheme II could be an excess in the high energy bins.
However it is clear that the sensitivity in the K2K experiment 
is not sufficient to either confirm or exclude one of the schemes.

In the planned CERN to Gran Sasso (CNGS) \cite{cgs} long baseline experiment, 
the main purpose will be to detect the appearance of 
$\nu_\mu \rightarrow \nu_\tau$. If experimental evidence of a 
$\nu_\tau$ appearance is found, a new and very important step in 
neutrino oscillation experiments will be made. A nearly 
pure $\nu_\mu$ beam will travel 732 km from CERN to the 
ICARUS \cite{icarus} and OPERA \cite{opera} experiments at Gran Sasso, 
with a mean energy of 17 GeV. 
The mean energy of neutrinos chosen in this experiment is high
so as to cross the $\tau$ production threshold and have
a sufficient $\tau$ production cross section.
This large value of the energy also turns out to be suitable for 
distinguishing between the two schemes, since 
the value of $\sin^2(\dmsq L/4E)$ is very small
for $\dmsq_{\rm atm}$ or $\dmsq_\odot$, but is oscillating for
$\dmsq_{LSND}$. Hence if the LSND angle is fairly large it can result 
in substantial contributions.

\begin{figure}[t,b,h]
\begin{center}
\mbox{
\epsfysize=7.3cm
\epsffile{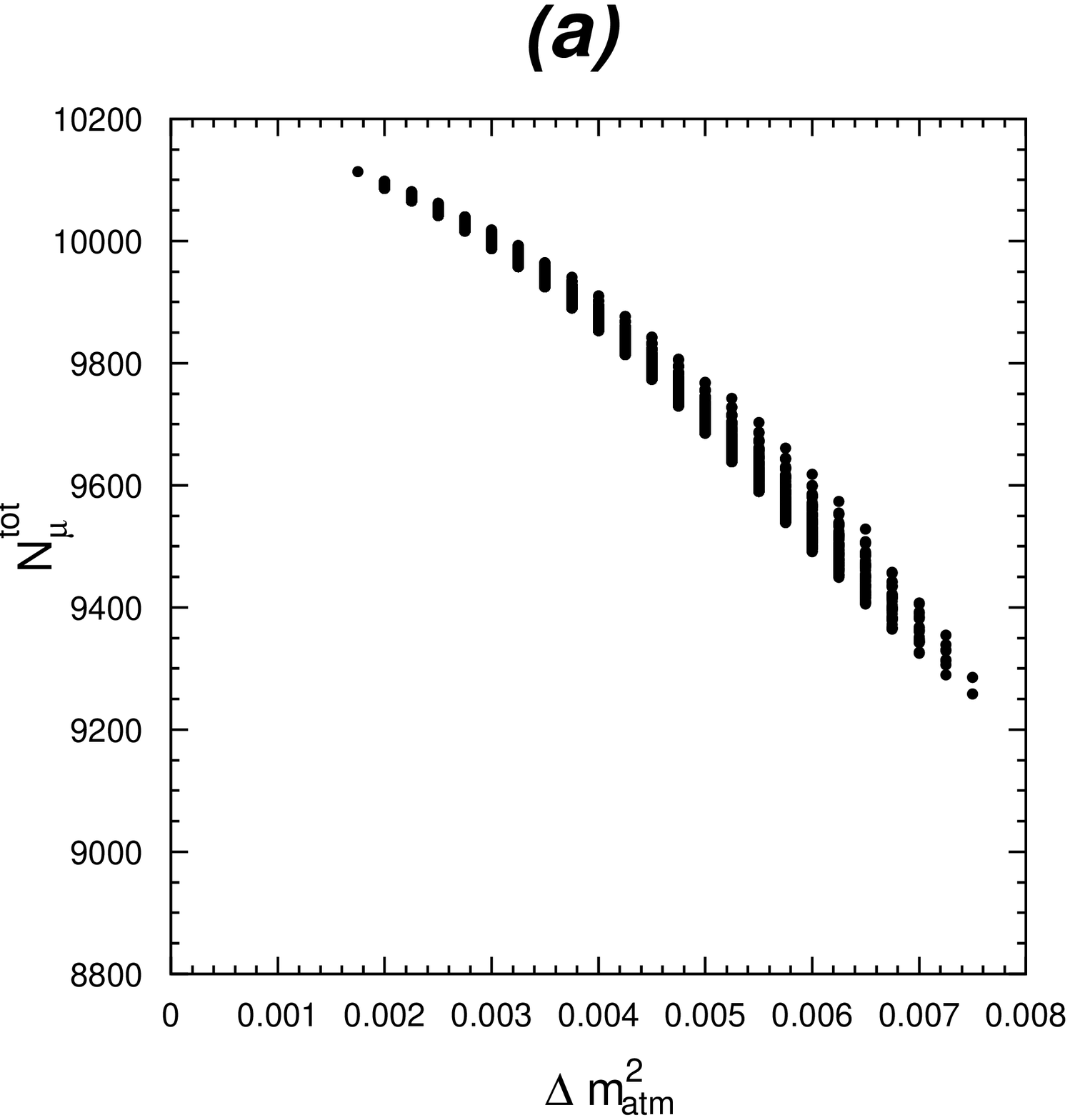}
}
\hspace{0.1cm}
\mbox{\vspace{1cm}}
\mbox{
\epsfysize=7.3cm
\epsffile{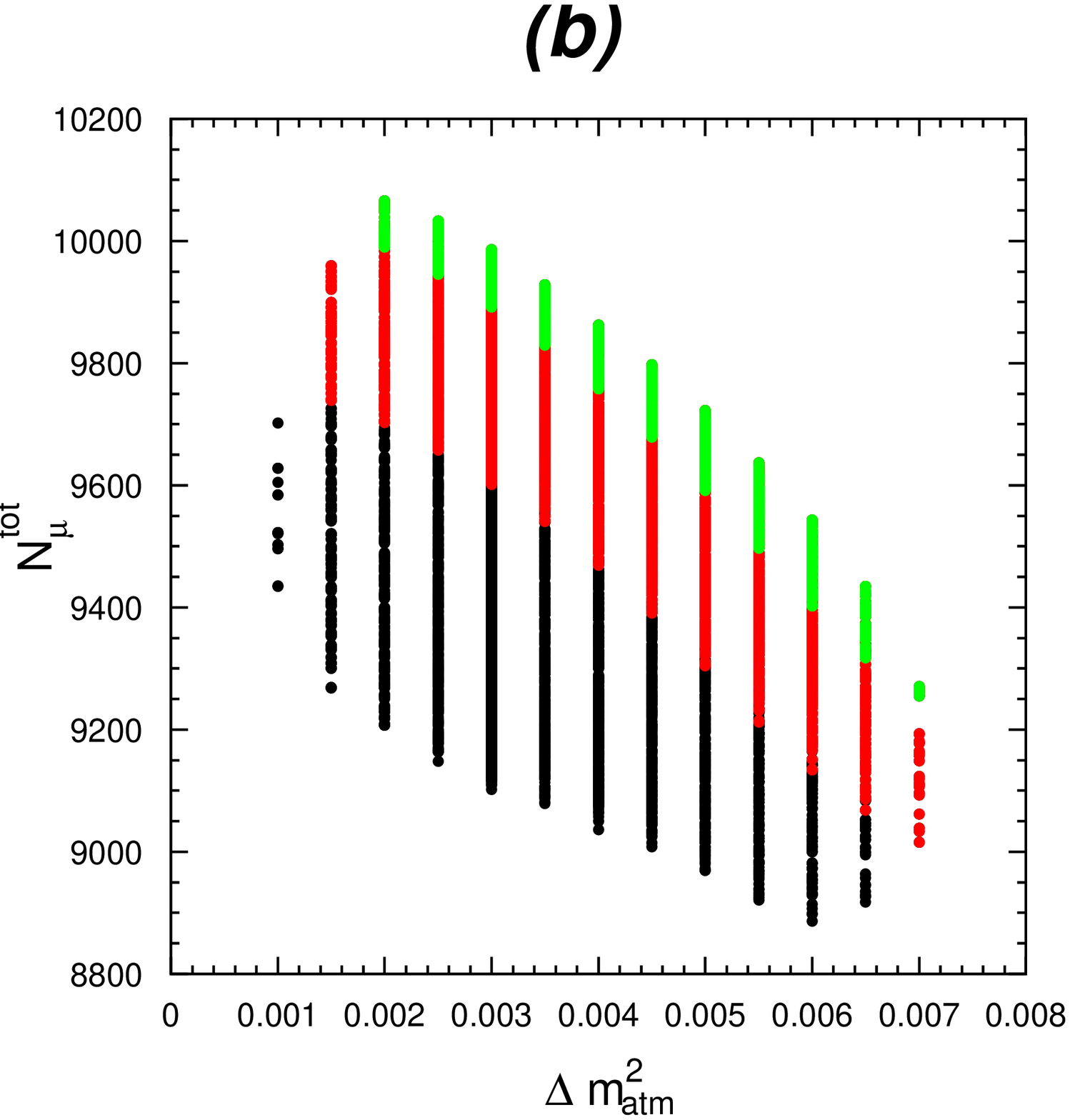}
}
\end{center}
\caption{The total number of $\nu_\mu$ CC events in CNGS calculated for all 
points within 99\% CL  
as a function of $\dmsq_{\rm atm}$ (a) for scheme I, 
(b) scheme II.  
The light gray/green points in (b) are with $\sin^2(2\theta_{\rm LSND})<0.02$ 
and the dark gray/red points are with $\sin^2(2\theta_{\rm LSND})<0.08$. 
The plot is for $2 \times 10^{25}$ pot or 5 years.} 
\label{totmu-cngs}
\end{figure}

We again calculate the total number of events in both $\nu_e$ and $\nu_\mu$ 
channels after $2 \cdot 10^{25}$ pot (corresponding to roughly 
five real years) and per kilo-ton according to (\ref{lbl}), 
for all points within 99\% CL.   
The measurement of $\nu_{\mu}$ (see Fig.~\ref{totmu-cngs}) will not reveal 
any difference between the two schemes unless $\dmsq_{\rm atm}$ is 
well known. 
Or, turned around, the accuracy of $\dmsq_{\rm atm}$ obtained 
in the case of scheme II is much weaker than that obtained in
the case of scheme I.

A large LSND angle will result in more $\nu_e$ appearance. 
For OPERA the expected sensitivity in the case of a negative search is 
$\sin^22\theta_{\mu e} \sim 1.5 \times 10^{-2}$ \cite{opera} for 
large $\D m^2$.  
The proposed sensitivity of the ICARUS detectors for large $\D m^2$ is  
$\sin^2(2\theta)>2.7 \times \;10^{-3}$. This is close to the limit 
obtained in Sec.~\ref{atm-lsnd} thereby testing nearly the whole LSND 
region, if the decision to build the detector is made.

\begin{figure}[t,b,h]
\begin{center}
\mbox{
\epsfysize=7.3cm
\epsffile{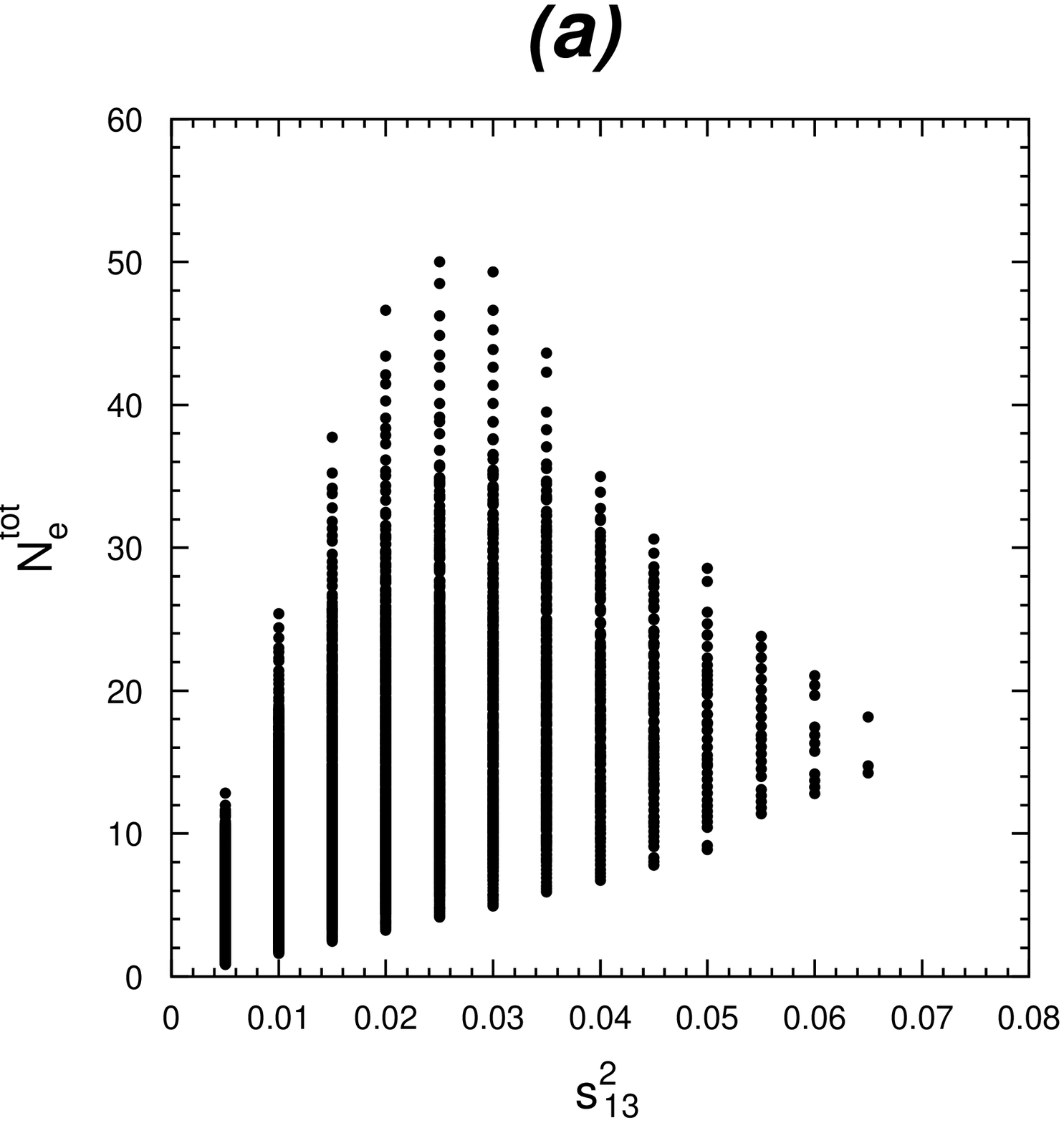}
}
\hspace{0.5cm}
\mbox{
\epsfysize=7.3cm
\epsffile{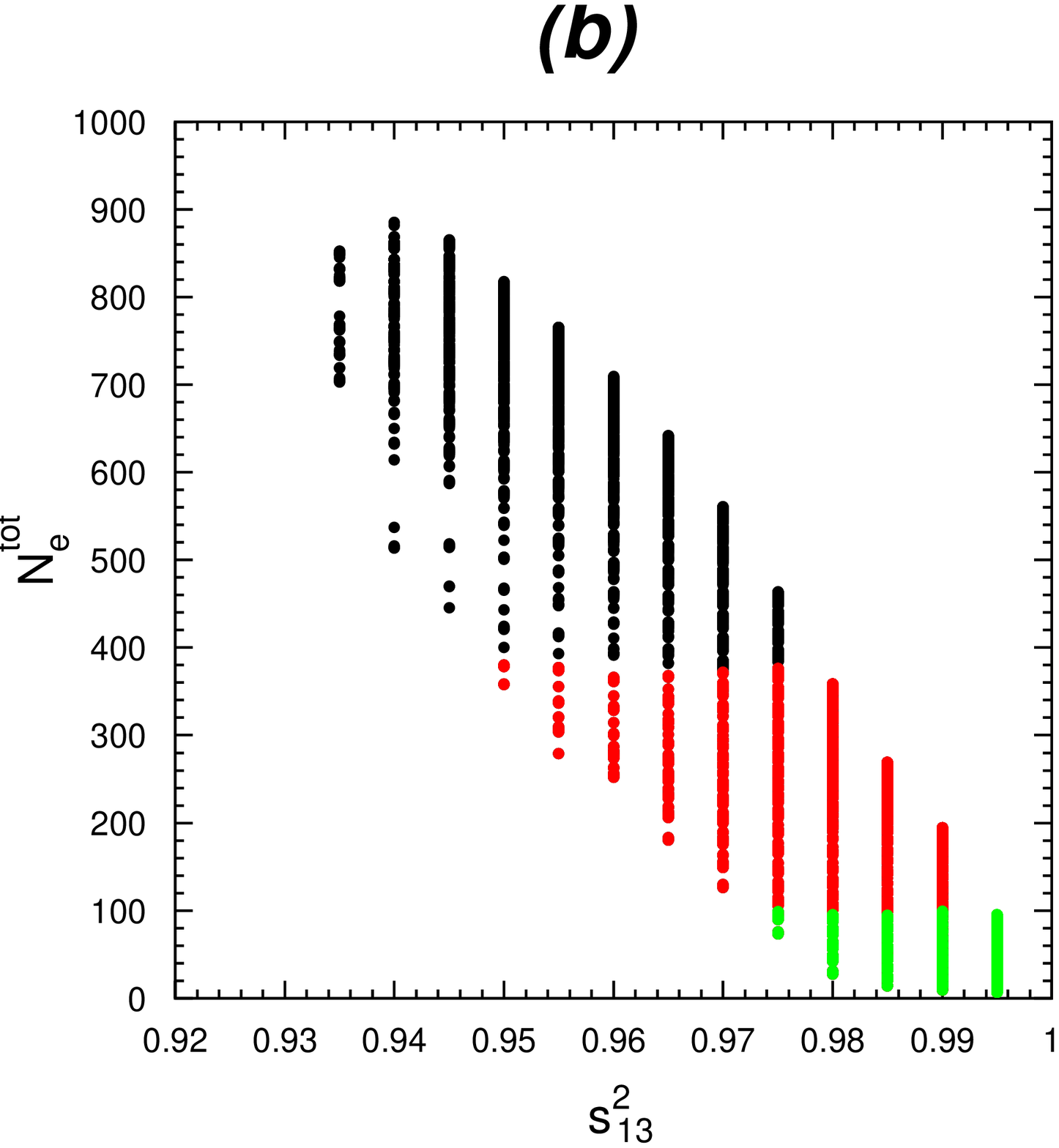}
}
\end{center}
\caption{The total number of $\nu_e$ CC events in CNGS calculated for 
all points within 99\% CL 
as a function $s_{13}^2$ (a) for scheme I, (b) for scheme II. 
The light gray/green points in (b) are with $\sin^2(2\theta_{\rm LSND})<0.02$ 
and the dark gray/red points are with $\sin^2(2\theta_{\rm LSND})<0.08$. 
The plot is for $2 \times 10^{25}$ pot or roughly 5 years.} 
\label{etot-cngs}
\end{figure}

\begin{figure}[h]
\begin{center}
\mbox{
\epsfysize=10cm
\input{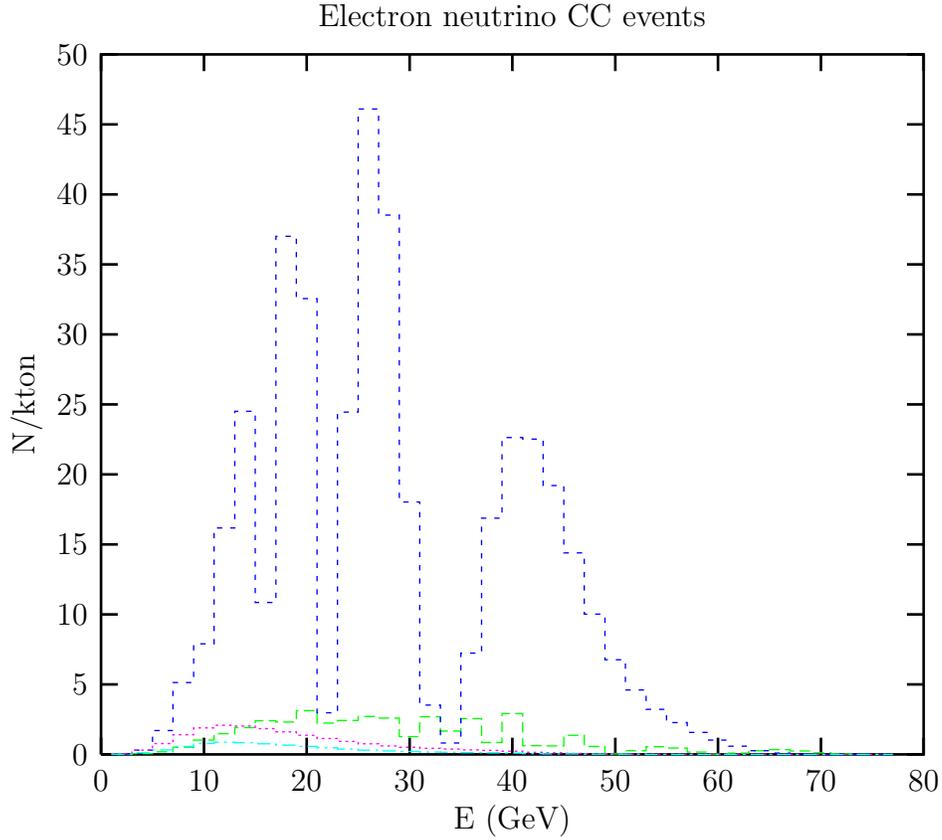}
}
\end{center}
\begin{center}
\begin{tabular}{|l||l|l|l|l|l|} \hline
Line type & $s_{12}^2$  & $s_{23}^2$ & $s_{13}^2$ & $\D m_{31}^2$ 
& $\D m_{21}^2$  \\ \hline 
Long-dashed & 0.10 & 0.20 & 0.99 & 1.0 & 0.003 \\ \hline
Short-Dashed & 0.98 & 0.74 & 0.97 & 0.22 & 0.003 \\ \hline
Dotted & 0.3 & 0.4 & 0.03 & 0.004 & 0.0002 \\ \hline
Dot-dashed & 0.3 & 0.4 & 0.01 & 0.004 & 0.0002 \\ \hline
\end{tabular} 
\end{center}
\caption{The expected number of electron CC events in the 
CERN to GRAN Sasso experiment as a function of the neutrino energy. The plot 
is for $2 \times 10^{25}$ pot or roughly 5 years.}
\label{cgsfig}
\end{figure}

The total number of electron neutrinos expected at CNGS as a function of
$s_{13}^2$ for both schemes is shown in Fig.~\ref{etot-cngs}.
As can be seen, the conventional scheme I
predicts an upper bound of $< 50$ on the number of electron
events observed, whereas in scheme II the number of
electron events can go as high as 400 for 
$\sin^2(2\theta_{\rm LSND})=0.08$. 
Scheme II can thus be 
identified if the number of events is observed to be large,
although it cannot be excluded on the basis of a low number of
events observed. It is evident that the two schemes will have quite 
different predictions, as the LSND scheme will predict, in most of the 
allowed parameter space, a larger number of electron events than the 
conventional one. Also the energy spectrum of $\nu_e$ events can be seen in 
Fig.~\ref{cgsfig}, using the same four points as in the K2K case above are 
shown. The energy dependence is also quite different. 
Hence we find that the CNGS experiments 
should be able to characterize the mass pattern. 

The MINOS experiment \cite{minos} will have features quite similar to 
those of the CNGS experiments. The baseline is the same, 
but the mean energy is lower (around 2 GeV). Therefore it will be 
harder to see any difference between the two schemes, again because of 
the dominance of the atmospheric mass squared oscillation. 
The total number of electron neutrino events is expected to be similar, 
with a maximum of 20 for scheme I and 28 for scheme II 
(calculated per kilo-ton and per year).  
In Fig. \ref{minosfig} we plot the energy spectrum of 
the electron neutrino events with a pure $\nu_\mu$ beam.  
It is seen that only in the case of a 
very large LSND angle will the experiment be able to get signs for 
an LSND mass squared difference. 
In order for MINOS to be able to distinguish between the schemes, the
medium or high energy option needs to be employed. Nevertheless the fact 
that the oscillation driven by the atmospheric mass squared difference is 
dominant allows an accurate determination of $\dmsq_{\rm atm}$, contrary 
to the CNGS experiments. 

\section{Summary}
\label{concl}

We explore the three-neutrino mixing scheme for solving the 
atmospheric and LSND anomalies, taking into account the constraints 
from CHOOZ. If the solar neutrino anomaly can be accounted for
by some exotic mechanism, this scheme can explain all the
observed neutrino experiments.

In order to check how well the atmospheric and LSND data can be
explained by a three-neutrino mixing scheme, we construct a
$\chi^2$ function that takes into account the sub-GeV and multi-GeV
electron and muon events observed at SK, the 
$\bar{\nu}_\mu \to \bar{\nu}_e$ conversion probability observed
at LSND, and the $\nu_e$ survival probability observed at CHOOZ.
We obtain a good fit, with $\chi^2/{\rm dof} \approx 1$. In order
to compare this goodness of fit with that of the conventional fit to
the solar, atmospheric and CHOOZ data, 
we also construct a $\chi^2$ function for the 
conventional scheme (I). We reproduce the features of the standard
three-neutrino fits, with $\chi^2/{\rm dof} \approx 1$. The
two fits are thus equally good: the ``goodness of fit'' 
criterion, as quantified by $\chi^2/{\rm dof}$, 
by itself does not favour either of these two schemes.
It is therefore necessary to investigate the often ignored
scheme (II) which accounts for the atmospheric, LSND and CHOOZ data
with three-neutrino oscillations.

We note some salient features of scheme II. 
The three-neutrino oscillation does provide a modest improvement of 
the fit with respect to the two-neutrino schemes. Large values of the 
LSND angles are favoured, with $\sin^2(2\theta_{\rm LSND}) \simeq 0.05$, 
even when including the Bugey data.  
There are almost no matter effects in this scheme, since the values of
the $\Delta m^2$'s are too large for the Earth's densities to have any effect.

The two schemes differ in certain respects, which may be exploited
for distinguishing between them. The ratios observed in the
atmospheric neutrinos, for example, should display different
behaviour. A clear signal for scheme II would be an excess of 
down-going multi-GeV electron neutrinos, accompanied by a small 
deficit for the down-going multi-GeV muon neutrinos. Also the sign 
of the up/down asymmetry for electron neutrino ratios is 
negative in the favoured region of parameter space, whereas it is positive 
in the conventional scheme. The current data, however, are 
insufficient to pick out one scheme over the other.

We also investigate the capability of the long baseline experiments
--- K2K, MINOS, CNGS --- to distinguish between the two schemes. 
We compute the $\nu_e$ and $\nu_\mu$ spectra
at these three experiments and find that for K2K and MINOS, the
$L/E$ value is too large and the statistics too small to observe 
any appreciable difference.
However, the final spectra at CNGS predicted by these two 
schemes are very different. 
We find that the observation of the final $\nu_\mu$ spectrum,
as well as the $\nu_\tau$ appearance events that may be observed
at CNGS, will not give us much useful information regarding the
choice of the scheme. The apparance of $\nu_e$, however, can show
some differences in principle, being large in scheme II. 
The observation of a large number of $\nu_e$ events
can rule out the conventional scheme, although a low
number of events cannot exclude scheme II.

Scheme II often gets a negatively biased treatment
compared with the conventional one, mainly because
it tries to explain the LSND results, which are not confirmed yet
by any other experiment. 
As we have shown in this paper, the goodness of fit cannot 
be reason to prefer one scheme over the other.
The ultimate arbiter of the issue is of course the experiment.
The BooNE experiment will be testing the whole LSND region.
With MiniBooNE proposed to be fully operative by fall 2002, it
will be the first to confirm or refute the LSND results. If the
LSND result is confirmed, the other experimental observations 
will have to be interpreted in terms of scheme II.

\vskip 0.5cm 
\noindent 
{\bf Acknowledgement}\\ 
The authors thanks D. Harris and M. Messier for information on MINOS and 
K. Nakamura for information on K2K. We are indebted to P. Hernandez and 
A. Romanino for useful conversations. 
S.S. is grateful to C. Pe\~{n}a-Garay for valuable discussions. 
S.S. would like to thank the CERN theory group, 
where most of this work was done, for their kind hospitality.
The work of S.S. was supported by NorFA, under the reference number
99.30.131-O.
\vskip 0.5cm     



\newpage
\noindent


\begin{figure}[t,b,h]
\vspace{-3.0cm}
\begin{center}
\mbox{
\epsfysize=8cm
\input{k2ke.eps}
}
\mbox{
\epsfysize=8cm
\input{k2kmu.eps}
}
\end{center}
\vspace{-0.3cm}
\begin{center}
\begin{tabular}{|l||l|l|l|l|l|} 
\hline
Line type & $s_{12}^2$  & $s_{23}^2$ & $s_{13}^2$ & $\D m_{31}^2$ & 
$\D m_{21}^2$  \\ \hline
Long-dashed & 0.10 & 0.20 & 0.99 & 1.0 & 0.003 \\ \hline
Short-Dashed & 0.98 & 0.74 & 0.97 & 0.22 & 0.003 \\ \hline
Dotted & 0.3 & 0.4 & 0.03 & 0.004 & 0.0002 \\ \hline
Dot-dashed & 0.3 & 0.4 & 0.01 & 0.004 & 0.0002 \\ \hline
\end{tabular} 
\end{center}
\vspace{-0.3cm}
\caption{The expected number of electron and muon events in the 
K2K experiment as a function of the neutrino energy. The  
solid curve corresponds to no oscillations in the muon event plot. 
The plot is for $10^{20}$ pot or roughly 3 years.}
\label{k2kfig}
\end{figure}


\begin{figure}[h]
\begin{center}
\mbox{
\epsfysize=10cm
\input{mine.eps}
}
\end{center}
\begin{center}
\begin{tabular}{|l||l|l|l|l|l|} \hline
Line type & $s_{12}^2$  & $s_{23}^2$ & $s_{13}^2$ & $\D m_{31}^2$ 
& $\D m_{21}^2$  \\ \hline 
Long-dashed & 0.10 & 0.20 & 0.99 & 1.0 & 0.003 \\ \hline
Short-Dashed & 0.98 & 0.74 & 0.97 & 0.22 & 0.003 \\ \hline
Dotted & 0.3 & 0.4 & 0.03 & 0.004 & 0.0002 \\ \hline
Dot-dashed & 0.3 & 0.4 & 0.01 & 0.004 & 0.0002 \\ \hline
\end{tabular} 
\end{center}
\caption{The expected number of electron CC events in the 
MINOS experiment as a function of the neutrino energy. }
\label{minosfig}
\end{figure}

\end{document}